\documentclass[11pt]{article}

\usepackage[utf8]{inputenc}
\usepackage[margin=1.3in]{geometry}
\usepackage{amsmath}
\usepackage{hyperref}
\usepackage{xcolor}
\usepackage{algorithm}
\usepackage{booktabs}
\usepackage{lscape}
\usepackage{algpseudocode}
\usepackage{etoolbox}
\usepackage{graphicx}
\usepackage{subcaption}

\usepackage[cmintegrals,cmbraces,ebgaramond]{newtxmath}
\usepackage{ebgaramond}

\usepackage[T1]{fontenc}

\newcommand{\orcid}[1]{\href{https://orcid.org/#1}{\includegraphics{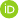}}}

\hypersetup{colorlinks,
            linkcolor={blue!80!black},
            citecolor={blue!80!black},
            urlcolor={blue!80!black}}

\title{MARLEM: A Multi-Agent Reinforcement Learning Simulation Framework for Implicit Cooperation in Decentralized Local Energy Markets}

\author{Nelson Salazar-Peña \orcid{0009-0003-7636-4150}\textsuperscript{a}, Alejandra Tabares \orcid{0000-0002-6630-5582}\textsuperscript{b}, and Andrés González-Mancera \orcid{0000-0002-0663-2653}\textsuperscript{a,}\footnote{Corresponding author\\ \textbf{Email addresses:} na.salazar10@uniandes.edu.co (Nelson Salazar-Peña\orcid{0009-0003-7636-4150}), a.tabaresp@uniandes.edu.co
(Alejandra Tabares\orcid{0000-0002-6630-5582}), angonzal@uniandes.edu.co (Andrés González-Mancera\orcid{0000-0002-0663-2653})
}}

\date{\small{
    \textsuperscript{a} Department of Mechanical Engineering, Universidad de los Andes, 111711, Bogotá D.C., Colombia\\
    \textsuperscript{b} Department of Industrial Engineering, Universidad de los Andes, 111711, Bogotá D.C., Colombia
}}

\begin{document}

\maketitle

\section*{Abstract}

The proliferation of distributed energy resources requires advanced coordination mechanisms for Local Energy Markets (LEMs) to maintain grid stability and economic efficiency, a challenge ill-suited to traditional centralized control. A significant research gap exists due to the lack of simulation frameworks that cohesively integrate realistic, decentralized market dynamics, physical grid constraints, and advanced Multi-Agent Reinforcement Learning (MARL) capabilities, especially for studying emergent coordination under partial observability. This paper introduces a novel, open-source MARL simulation framework for studying implicit cooperation in LEMs, modeled as a decentralized partially observable Markov decision process and implemented as a Gymnasium environment for MARL. Our framework features a modular market platform with plug-and-play clearing mechanisms, physically constrained agent models (including battery storage), a realistic grid network, and a comprehensive analytics suite to evaluate emergent coordination. The main contribution is a novel method to foster implicit cooperation, where agents' observations and rewards are enhanced with system-level key performance indicators to enable them to independently learn strategies that benefit the entire system and aim for collectively beneficial outcomes without explicit communication. Through representative case studies (available in a dedicated GitHub repository in \href{https://github.com/salazarna/marlem}{https://github.com/salazarna/marlem}), we show the framework's ability to analyze how different market configurations (such as varying storage deployment) impact system performance. This illustrates its potential to facilitate emergent coordination, improve market efficiency, and strengthen grid stability. The proposed simulation framework is a flexible, extensible, and reproducible tool for researchers and practitioners to design, test, and validate strategies for future intelligent, decentralized energy systems. \newline

\footnotesize{\textbf{\textit{Keywords}} \hspace{2mm} Local Energy Markets, Multi-Agent Reinforcement Learning, Decentralized Partially Observable Markov Decision Process, Implicit Cooperation, Decentralized Energy Systems, Simulation Framework}

\section{Introduction}
\label{1-sec:1-introduction}

\subsection{Motivation}
\label{1-sec:1.1-motivation}

The global energy sector is undergoing a fundamental paradigm shift, transitioning from a historically centralized generation model, reliant on a few large-scale, dispatchable fossil-fuel power plants, towards a highly decentralized system characterized by the extensive proliferation of Distributed Energy Resources (DERs) \cite{feng2024, weidlich2018}. This category of assets includes residential and commercial rooftop solar photovoltaics, battery energy storage systems, electric vehicles, and controllable loads \cite{gridworks2017coordination}. This transformation is driven by a confluence of factors: international climate policy mandates for decarbonization, the declining costs of renewable and storage technologies making them economically viable for consumers, and a rising demand for energy autonomy and resilience against large-scale grid disruptions \cite{weidlich2018}.

This evolution marks the rise of the prosumer, an active market participant who not only consumes energy but also produces, stores, and potentially manages it \cite{qiu2021}. While this paradigm shift presents significant opportunities for a more efficient, resilient, and sustainable energy future, it concurrently introduces operational challenges, particularly for distribution networks \cite{harder2025}. The traditional unidirectional flow of power from central generators to passive consumers is being replaced by complex, bidirectional flows involving numerous, heterogeneous actors. This necessitates more sophisticated methods of coordination and control to manage potential issues like grid congestion, voltage instability, and the inherent variability of renewable generation \cite{harder2025}.

In this context, Local Energy Markets (LEMs) have emerged as a promising framework for managing this emergent complexity at the distribution level \cite{feng2024, weidlich2018}. By offering a platform for peer-to-peer (P2P) energy trading and flexibility services within a specific geographic community, LEMs aim to enhance local grid stability, promote the efficient utilization of local renewable resources, reduce reliance on bulk transmission systems (thereby minimizing losses), and unlock the full economic potential of distributed assets by allowing prosumers to directly monetize their flexibility \cite{feng2024, weidlich2018}.

\subsection{Problem statement}
\label{1-sec:1.2-problem-statement}

Despite their potential, the effective design and operation of efficient, stable, and scalable LEMs present a complex scientific and engineering challenge, often conceptualized as a trilemma \cite{feng2024, weidlich2018}. First, achieving efficient and scalable coordination among a potentially massive population of autonomous, self-interested agents is a problem of immense complexity \cite{qiu2023}. Each agent (representing a home, a building, an electric vehicle, or a community battery) makes decisions based on private objectives (e.g., minimizing costs, maximizing revenue) and limited, local information. This inherent decentralization creates a non-stationary environment where the optimal strategy for any single agent depends on the concurrent, often unobservable, actions of all other participants \cite{weidlich2018, henry2021}.

Second, the economic transactions constituting the market must not compromise the physical integrity of the underlying distribution network \cite{feng2024}. Energy trades correspond to physical power injections and withdrawals. Uncoordinated actions, even if economically rational for individual agents (e.g., simultaneous battery discharging during peak prices), can lead to localized voltage violations, thermal overloading of lines, or unacceptable power quality degradation, threatening grid security \cite{feng2024, weidlich2018}. Therefore, any viable LEM design must inherently respect or be co-managed with these physical constraints. LEM simulation software must thus tightly couple economic decision-making with physical grid simulation to capture these techno-economic interactions.

Third, a practical LEM architecture must preserve the privacy and autonomy of its participants \cite{qiu2023}. Centralized control solutions, which require agents to divulge sensitive consumption data or cede control authority to a central entity, often face significant barriers due to privacy concerns, cybersecurity risks, and the potential creation of single points of failure \cite{weidlich2018, qiu2023}. Scalability also becomes a major issue, as a central controller managing potentially a large number (hundreds or thousands) of DERs faces immense computational burdens. Thus, truly decentralized approaches that respect agent autonomy and data privacy are highly desirable, if not essential, necessitating software capable of modeling and rigorously evaluating such paradigms.

The central problem, therefore, lies in discovering mechanisms and system designs, and the appropriate simulation software to develop and test them, through which desirable system-level goals (supply-demand balance, aggregate economic efficiency, physical grid security) can emerge from the uncoordinated, self-interested actions of independent entities operating under partial observability, without resorting to untenable centralized control.

\subsection{Research gap}
\label{1-sec:1.3-research-gap}

Addressing the LEM trilemma requires sophisticated modeling and simulation software capable of capturing the intricate interplay between adaptive agent learning (particularly multi-agent reinforcement learning (MARL)), complex market dynamics, and realistic physical grid constraints. However, a comprehensive review of the existing literature (detailed in Section \ref{1-sec:2-literature-review}) reveals a significant and critical gap: there is no existing software that effectively unifies realistic electrical network simulation, flexible energy market modeling, and agent-based MARL within a single, standardized, easy-to-use framework \cite{charbonnier2023}.

Existing simulation tools often fall short due to fragmentation and specialization, hindering progress in understanding holistically integrated LEMs \cite{weidlich2018, henry2021, charbonnier2023}:

\begin{itemize}
    \item \textbf{Fragmentation:} Many traditional power system simulators (e.g., GridLAB-D, MATPOWER) excel at physical modeling but lack native support for agent-based market interactions and adaptive learning capabilities \cite{czekster2020, chassin2008, zimmerman2011}. Conversely, agent-based modeling platforms tailored for energy markets (e.g., Lemlab/Hamlet \cite{zade2022}) often focus on optimization-based agents or simplified heuristics, lacking robust integration with state-of-the-art MARL algorithms needed to study emergent adaptive strategies \cite{harder2025, mbungu2019}. This forces researchers into complex co-simulation setups or requires them to abstract crucial aspects of the system.

    \item \textbf{Abstraction:} Numerous platforms, particularly those focused on specific RL applications like demand response (e.g., CityLearn \cite{harder2025} employs "copper-plate" grid models, ignoring critical physical constraints like congestion and losses. This severely limits research vital techno-economic interactions and the potential for market actions to impact grid stability \cite{charbonnier2023}. Others abstract the market mechanism itself, preventing the study of how different auction types or pricing rules influence agent behavior \cite{henry2021}.

    \item \textbf{Centralization bias:} Frameworks integrating RL often target wholesale markets (e.g., ASSUME \cite{harder2025} or inherently rely on Centralized Training with Decentralized Execution (CTDE) paradigms \cite{harder2025, vantilburg2023}. While CTDE is practical, it still necessitates significant centralized coordination during the learning phase, fundamentally contradicting the principles of truly decentralized, private, and resilient LEMs, and failing to capture the unique challenges of fully decentralized learning \cite{charbonnier2023}.

    \item \textbf{Limited focus on implicit coordination:} Critically, no existing standardized, extensible, and MARL-native simulation framework is specifically designed to investigate implicit coordination, where agents learn to coordinate without explicit communication by responding to shared environmental signals, within a physically constrained, decentralized market under the challenging Decentralized Training, Decentralized Execution (DTDE) paradigm \cite{weidlich2018, charbonnier2023}. This capability is essential for exploring scalable and privacy-preserving coordination mechanisms.
\end{itemize}

This lack of integrated, standardized, and appropriately focused tooling significantly hinders progress in understanding, designing, and deploying truly decentralized, intelligent, and physically-aware LEMs.

\subsection{Novelty and contribution}
\label{1-sec:1.4-contribution}

This research introduces a novel, open-source simulation framework engineered for investigating MARL in decentralized LEMs. This software is designed to fill the identified gap by providing the first platform that cohesively integrates the necessary components for this research area. Specifically, this work addresses the core challenges of LEM design and operation by enabling studies into:

\begin{itemize}
    \item How multi-agent simulation environments, particularly their information and incentive structures (observation and reward design), can be engineered within our unified software to foster the emergence of implicit cooperation among self-interested MARL agents, thereby enhancing grid stability and economic efficiency.

    \item How different decentralized market mechanisms (e.g., auction types, pricing rules), implemented modularly in the software, influence the learning dynamics, strategic behavior of agents, and the resultant market-level coordination and economic outcomes within a unified simulation context.

    \item The impact of explicitly modeled physical grid constraints, such as transmission losses and congestion (simulated via the software's integrated grid component), on the trading strategies adopted by MARL agents and on overall market performance and stability.
\end{itemize}

The presented LEM software framework offers several contributions and novelties specifically aimed at bridging the identified gaps in existing tooling:

\begin{enumerate}
    \item \textbf{Unified techno-economic MARL environment.} It uniquely combines a modular, MARL-compatible market (including novel preference-based matching) with a physically realistic distribution grid model (including losses and congestion) within a single, integrated Gymnasium environment, explicitly overcoming the market-vs-grid dichotomy and fragmentation prevalent in prior tools \cite{charbonnier2023}.

    \item \textbf{Focus on full decentralization.} Unlike many existing tools biased towards CTDE \cite{harder2025, nweye2025}, our software is architected to natively support and facilitate research into the DTDE paradigm \cite{wen2021, dominguez-garcia}, aligning with the principles and requirements of real-world decentralized systems \cite{charbonnier2023}.

    \item \textbf{Implicit cooperation as a core construct.} Its observation and reward structures are explicitly engineered using system-level key performance indicators (KPIs) to enable the study of emergent cooperation driven by shared environmental signals \cite{lia, li2021}, moving beyond purely profit-maximizing objectives towards system-aware agent learning within a fully decentralized context \cite{weidlich2018, charbonnier2023}.

    \item \textbf{Standardization and modularity.} Adherence to the widely adopted Gymnasium standard \cite{vazquez-canteli} ensures interoperability with a vast ecosystem of RL algorithms, while its modular design (especially for market and grid components) promotes reproducibility, extensibility, and ease of use for the broader research community \cite{weidlich2018, henry2021, charbonnier2023}.

    \item \textbf{Comprehensive analytics suite.} It includes a set of integrated tools for quantifying emergent behaviors, market performance, grid stability, and coordination effectiveness, providing the empirical tools necessary to validate scientific claims directly within the software \cite{henry2021}.
\end{enumerate}

By providing this unique combination of features within a single, accessible software package, our framework serves as a vital tool for advancing the understanding and development of truly decentralized, intelligent, physically-aware, and efficient LEMs capable of supporting the future energy grid \cite{charbonnier2023}.

\subsection{Paper structure}
\label{1-sec:1.5-paper-structure}

This paper provides a detailed exposition of the framework's design, implementation, and capabilities. The remainder of the manuscript is structured as follows: Section \ref{1-sec:2-literature-review} provides a comprehensive review of related work, further detailing the research gap our framework addresses. Section \ref{1-sec:3-methodology} offers a detailed description of the framework's methodology and architecture, covering the Dec-POMDP formulation, agent models, market design, grid integration, implicit cooperation mechanisms, and MARL paradigms. Section \ref{1-sec:4-experimental-setup} outlines the experimental setup used to demonstrate the framework's capabilities. Section \ref{1-sec:5-results-and-discussion} presents and discusses the results from this experiment. Finally, Section \ref{1-sec:6-conclusion} concludes the paper, summarizes the key contributions, and delineates directions for future research.

\section{Literature review}
\label{1-sec:2-literature-review}

The development of our simulation framework is situated at the confluence of evolving research domains: decentralized energy systems, multi-agent systems, MARL for power system management, and the design of specialized simulation environments. A thorough review of the state-of-the-art across these areas is crucial to contextualize our contributions and delineate the research gap that our work aims to address. This section undertakes such a review, analyzing existing methodologies, algorithms, and simulation tools relevant to LEMs.

\subsection{Centralized and decentralized energy markets}
\label{1-sec:2.1-centralized-and-decentralized}

The traditional paradigm for power grid management is inherently centralized, relying on a single entity, such as an Independent System Operator, to solve large-scale optimization problems for unit commitment and economic dispatch. This top-down approach has historically ensured grid reliability. However, the escalating penetration of geographically dispersed and intermittent DERs fundamentally challenges the scalability and efficacy of centralized control. The significant volume of data and the high-frequency decision-making required to coordinate hundreds or thousands of DERs render traditional methods computationally intractable and inefficient. 

In response, a significant research trend has emerged toward decentralized energy systems, particularly in the form of LEMs. LEMs propose a bottom-up approach where prosumers can trade energy directly with one another in a P2P fashion. The theoretical advantages are numerous: enhanced grid resilience by reducing reliance on long-distance transmission, improved economic efficiency by minimizing transaction costs, and greater prosumer empowerment. However, these benefits are contingent upon solving the complex coordination problem at the local level. The primary trade-off is between the theoretical optimality of a centralized controller with perfect information and the scalability, robustness, and privacy afforded by a decentralized system. Our work is positioned in the latter area, investigating mechanisms that can close this performance gap by enabling effective coordination without a central authority.

\subsection{Multi-agent systems in energy grid management}
\label{1-sec:2.2-mas-energy-management}

The conceptualization of power systems as multi-agent systems is a well-established field, offering a natural paradigm for modeling the interactions of numerous distributed entities \cite{keren, biagioni2022}. Early applications primarily utilized rule-based or optimization-based agents to tackle challenges such as load shedding, fault diagnosis, power restoration, and microgrid control \cite{zhu2024}. These foundational works demonstrated the feasibility of decentralized coordination using predefined communication protocols and heuristic strategies, often focused on achieving system stability or specific operational objectives \cite{czekster2020, keren}. For instance, multi-agent systems have been employed for resource allocation and scheduling \cite{zhu2024} and coordinating Transmission System Operators and Distribution System Operators (DSOs) \cite{givisiez2020, alazemi2022}. The strength of multi-agent systems lies in its ability to model heterogeneity, bounded rationality, and emergent phenomena inherent in complex socio-technical systems like the power grid \cite{klein2019, trimarchi2025}.

However, these traditional approaches to multi-agent systems often assume agents follow fixed rules or possess the computational capacity to solve complex local optimization problems, which may be unrealistic for heterogeneous prosumers \cite{harder2025, klein2019}. More critically, they typically lack the capacity for adaptive learning in dynamic, uncertain environments; agent behaviors are pre-programmed and cannot evolve in response to changing market conditions or the strategic adaptations of other participants \cite{klein2019}. While valuable for demonstrating the potential of distributed control, these static approaches fall short of capturing the adaptive, strategic learning required in modern, competitive LEMs where emergent behaviors and complex interactions dominate \cite{vantilburg2023, akhatova2022}.

\subsection{Application and limitations of MARL for DER coordination}
\label{1-sec:2.3-marl-der-coordination}

MARL has emerged as a particularly promising paradigm for overcoming the limitations of traditional multi-agent systems, enabling autonomous agents to learn sophisticated, adaptive strategies directly from interaction within complex, non-stationary environments \cite{li2022structured, vazquez-canteli2019}. Its model-free nature allows agents to operate effectively even without explicit system models, making it well-suited for the inherent uncertainties of LEMs \cite{vazquez-canteli2019}. The application of MARL in energy systems spans several key areas:

\begin{itemize}
    \item \textbf{P2P energy trading:} This is a core application where MARL agents learn bidding and scheduling strategies to trade energy surpluses and deficits directly \cite{qiu2021, qiu2023, may2023, sabillon2021}. Studies vary significantly in market mechanism complexity, ranging from detailed double-sided auctions \cite{qiu2021, qiu2023} or continuous double auctions \cite{qiu2023} to highly simplified mechanisms based on supply-demand ratios \cite{feng2024} or abstracted clearing prices. While double auctions offer high economic fidelity, simplified models are often employed when integrating complex physical constraints \cite{feng2024}. Some approaches also utilize traditional distributed optimization methods like Alternating Direction Method of Multipliers as components or baselines \cite{may2023}, though these often struggle with practical issues and unrealistic agent assumptions \cite{feng2024}.

    \item \textbf{Ancillary services:} MARL is explored for coordinating DERs to provide essential grid support services, such as voltage control \cite{qiu} and frequency regulation \cite{feng2024, qiu}. These applications often focus on the physical control aspects, sometimes simplifying the economic incentives or market structures involved, treating the problem more as a cooperative control task than a market interaction.

    \item \textbf{Demand response:} MARL agents learn to adjust consumption patterns in response to price signals or direct requests from aggregators or utilities \cite{vantilburg2023, oh2022}. Studies like MARL-iDR \cite{vantilburg2023} demonstrate effective coordination for incentive-based demand response, often using CTDE paradigms \cite{vantilburg2023}.
\end{itemize}

Despite promising results, a critical analysis of the existing literature reveals common limitations in how the MARL problem is often formulated for LEMs:

\begin{itemize}
    \item \textbf{State space:} Many studies provide agents primarily with local state information (e.g., state of charge (SoC), profit) and basic market signals (e.g., previous clearing prices) \cite{qiu2021, may2023}. While some works incorporate local grid information (e.g., voltage levels \cite{feng2024}, the strategic use of system-level KPIs, such as overall grid congestion, social welfare, or market imbalance, as shared observational signals to facilitate implicit coordination without direct communication is significantly underdeveloped. Agents often lack the necessary contextual information to understand the broader system impact of their actions.

    \item \textbf{Action space:} The action space is frequently simplified. Many studies utilize discrete actions (e.g., charge/discharge levels \cite{biagioni2022} or limit continuous actions primarily to bidding quantities, assuming prices are fixed or determined centrally \cite{markgraf2023}. Frameworks modeling sophisticated P2P markets often lack explicit action components for strategic partner selection or reputation-based interactions, limiting the potential for emergent social dynamics. Our framework employs a continuous, multi-dimensional action space including price, quantity, buy/sell direction, and a preferred partner ID, enabling richer strategic learning that encompasses both economic and relational aspects.

    \item \textbf{Reward function:} The predominant approach relies heavily on maximizing individual agent profit \cite{qiu2021, qiu2023, may2023}. While economically intuitive, purely profit-driven rewards frequently lead to collectively undesirable outcomes, such as market instability, strategic withholding, or disregard for physical grid constraints \cite{feng2024, markgraf2023}. Although some works incorporate penalties for constraint violations (e.g., voltage penalties \cite{feng2024}), few employ structured, multi-objective rewards designed to explicitly incentivize emergent cooperation by directly linking individual rewards to positive contributions towards system-level KPIs. Our framework's reward function, featuring a base reward modulated by a multiplicative market-wide cooperation factor and an individual contribution factor, represents a novel attempt to systematically address this gap and promote pro-social learning by making collective success directly beneficial to the individual.
\end{itemize}

A challenge identified across these applications is a market-versus-grid dichotomy: studies excelling in market fidelity often abstract the grid as an unconstrained "copper plate" \cite{qiu2021, qiu2023, guerrero2018a}, while those with detailed grid models tend to simplify the market significantly \cite{feng2024, markgraf2023}. This trade-off hinders the development of holistically realistic LEM simulations capable of capturing the critical techno-economic interdependencies.

\subsection{Simulation frameworks for LEMs and MARL}
\label{1-sec:2.4-simulation-frameworks}

The evaluation and benchmarking of MARL strategies rely on suitable simulation environments. While numerous tools exist, a comparative review highlights the lack of a single platform adequately addressing the specific needs of research into decentralized, physically-constrained LEMs, particularly those focused on implicit coordination and fully decentralized learning.

\begin{itemize}
    \item \textbf{Agent-based modeling tools:}  Frameworks such as Lemlab/Hamlet \cite{zade2022} offer modular agent-based modeling architectures specifically for LEM research \cite{harder2025}. They provide tools for testing various market designs and clearing mechanisms but primarily focus on traditional economic modeling and optimization-based agents. They generally lack native, integrated support for MARL algorithms and training paradigms, making it difficult to study adaptive learning dynamics \cite{mbungu2019}.

    \item \textbf{Optimization-based tools:} Frameworks like OPLEM \cite{essayeh2024} and SIMTES \cite{melo2023} aim to integrate market design with network operation assessment \cite{harder2025}. OPLEM includes modules for various market types (P2P, auctions, time-of-use), while SIMTES employs co-simulation (integrating multi-agent systems, Pandapower \cite{thurner2018}, ns-3 \cite{wehrle2010}. Both incorporate detailed grid models. However, their core logic relies on mathematical optimization solvers (linear programming, mixed integer linear programming) for agent decisions and market clearing, rather than end-to-end MARL \cite{harder2025}. They serve well for benchmarking optimal solutions under simplifying assumptions but do not capture the learning dynamics or potential sub-optimalities inherent in MARL. SINTES appears to fall into this category as well, focusing on optimization within transactive energy systems.

    \item \textbf{Specialized RL environments for energy:}
    \begin{itemize}
        \item \textbf{CityLearn:} A widely adopted, Gymnasium-compliant environment focused on multi-agent demand response and building energy coordination \cite{harder2025, nweye2025, vazquez-canteli}. Its primary strength is standardizing demand response research. However, it lacks an explicit P2P market mechanism and abstracts away the physical power grid, making it unsuitable for studying market-grid interactions, P2P trading dynamics, or physical constraints like congestion \cite{harder2025, vazquez-canteli}.

        \item \textbf{ASSUME:} An agent-based framework modeling wholesale electricity markets with RL integration (specifically the multi-agent TD3 algorithm under a CTDE paradigm) \cite{harder2025, steinbrink2019}. It features highly modular market designs (including EUPHEMIA-like auctions) and detailed grid modeling (PyPSA integration) \cite{harder2025}. However, its focus on wholesale markets and its inherent reliance on CTDE make it unsuitable for studying fully decentralized LEMs at the distribution level \cite{harder2025}.

        \item \textbf{RL-ADN:} A high-performance RL environment focused on optimal battery storage dispatch in active distribution networks, featuring a "tensor power flow" solver \cite{hou2025}. It excels in grid control research but lacks market mechanisms entirely \cite{hou2025}.

        \item Other tools like energy-py \cite{energypy2024} focus on single-agent optimization (e.g., battery storage), while marl-local-electricity applies MARL but is not presented as a general-purpose, modular framework \cite{mbungu2019}.
    \end{itemize}

    \item \textbf{Power system simulators:} Tools like GridLAB-D (distribution networks) and MATPOWER (transmission networks) provide high-fidelity physical grid modeling but are not inherently agent-based market simulators and lack MARL integration \cite{czekster2020}. Grid2Op offers a powerful Gymnasium environment for transmission grid operation from a centralized operator perspective, unsuitable for decentralized market simulation \cite{marot2021}.

    \item \textbf{General MARL frameworks:} Libraries like PyMARL/EPyMARL \cite{samvelyan2019, papoudakis2021benchmarking} and MARLlib \cite{hu2022marllib} provide excellent algorithmic implementations but are domain-agnostic algorithm libraries, not environments. They often focus architecturally on CTDE paradigms and lack the built-in domain-specific energy models needed for LEM research. Co-simulation frameworks like Mosaik \cite{steinbrink2019} offer flexibility but require significant integration effort to combine disparate simulators (e.g., for market logic, grid physics, and agent learning) \cite{steinbrink2019, steinbrink2018}.
\end{itemize}

Table \ref{tab:simulation-frameworks} summarizes the key characteristics of the reviewed frameworks in relation to the requirements for studying decentralized, physically-aware LEMs using MARL, particularly focusing on DTDE and implicit coordination.

\begin{table}[h!]
\centering
\caption{Simulation frameworks for local energy markets and multi-agent reinforcement learning.}
\label{tab:simulation-frameworks}
\resizebox{\textwidth}{!}{%
\begin{tabular}{@{}lllllll@{}}
\toprule
\textbf{Framework} & \textbf{Primary Focus} & \textbf{Gym Compliant?} & \textbf{Modular Market?} & \textbf{Grid Detail} & \textbf{Supports DTDE?} & \textbf{Ref} \\ \midrule
Lemlab/Hamlet & LEM (Optimization) & No & Yes (Optimization-based) & Varies & N/A & \cite{zade2022} \\
OPLEM/SIMTES & LEM/TES (Optimization) & No & Yes (Optimization-based) & Detailed & N/A & \cite{essayeh2024, melo2023} \\
CityLearn & Demand Response & Yes & No & Abstracted & Execution only & \cite{nweye2025, vazquez-canteli} \\
ASSUME & Wholesale Markets (RL) & No & Yes (Auction-based) & Detailed & No (CTDE) & \cite{harder2025} \\
RL-ADN & ESS Dispatch (RL) & No & No & High & Not specified & \cite{hou2025} \\
Grid2Op & Transmission Grid Ops (RL) & Yes & No & High & Paradigm-agnostic & \cite{marot2021} \\
PyMARL/MARLlib & MARL Algorithms & N/A & N/A & N/A & Limited (CTDE) & \cite{samvelyan2019, papoudakis2021benchmarking, hu2022marllib} \\
\textbf{Our Framework} & \textbf{Decentralized LEMs (MARL)} & \textbf{Yes} & \textbf{Yes (MARL-native)} & \textbf{Detailed} & \textbf{Yes (Core Feature)} & - \\ \bottomrule
\end{tabular}%
}
\end{table}

\subsection{Research gap and contribution opportunity}
\label{1-sec:2.5-contribution-opportunity}

This literature review reveals significant gaps in the existing landscape of simulation tools and MARL applications for energy systems. The market-versus-grid dichotomy persists, with few frameworks successfully integrating high-fidelity models of both domains within a learning-based paradigm \cite{qiu2021, markgraf2023}. Furthermore, the centralization-versus-decentralization dilemma remains unresolved in MARL training approaches \cite{nweye2025, tian2020}. Most advanced MARL energy frameworks rely on CTDE \cite{harder2025, vantilburg2023}, which, while practical for training, requires a centralized entity with global information, fundamentally contradicting the principles of a truly decentralized, private, and resilient LEM \cite{marot2021, fonseca2025}.

Consequently, a clear research gap exists for a MARL-native simulation framework specifically designed for decentralized LEMs that simultaneously features (i) a modular market mechanism suitable for MARL agents, (ii) the incorporation of realistic physical grid constraints (losses, congestion), (iii) is built upon and facilitates research into the fully DTDE paradigm, and (iv) explicitly incorporates mechanisms (like shared KPI signals in observations and rewards) to study and foster implicit cooperation.

Our proposed framework is designed to fill this gap. It provides the research community with the first standardized, Gymnasium-compliant, open-source tool that cohesively integrates these four essential properties (recall Section \ref{1-sec:1.4-contribution}). By providing this combination, our framework serves as a tool for advancing the understanding and development of truly decentralized, intelligent, physically-aware, and efficient LEMs capable of supporting the future energy grid.

\section{Methodology}
\label{1-sec:3-methodology}

The simulation framework detailed in this paper is a comprehensive, open-source tool developed for the analysis of LEMs within a MARL paradigm. The problem is formally structured as a Dec-POMDP to accurately model the challenges of decision-making under uncertainty and limited information inherent in such decentralized systems. This formalization provides the mathematical foundation upon which the interactions between agents and their environment are built, ensuring a principled approach to modeling the complex dynamics of the LEM.

A core design principle of the framework is its implementation as a multi-agent environment conforming to the Gymnasium standard. This adherence is a critical contribution, as it ensures compatibility with an extensive range of state-of-the-art RL libraries and algorithms. By adopting this widely recognized interface, the framework lowers the barrier to entry for other researchers, facilitating reproducible experiments and allowing for the standardized benchmarking of new MARL strategies.

 This documentation is structured to guide the user from a high-level understanding to granular implementation details, and includes detailed descriptions for each core module, a step-by-step description of the simulation loop, detailing how the modules interact at each trading period, and specifics on the training and inference processes, including the MARL algorithms and training paradigms supported.

This detailed documentation ensures that the framework can serve not only as a tool for validation but also as a platform for future extension and development by the research community. For researchers seeking to engage with the framework at a deeper technical level, supplementary material is provided in a dedicated GitHub repository (\href{https://github.com/salazarna/marlem}{https://github.com/salazarna/marlem}).

\subsection{Simulation framework architecture}
\label{1-sec:3.1-framework-architecture}

\begin{figure}[h!]
    \centering
    \includegraphics[width=\textwidth, angle=0]{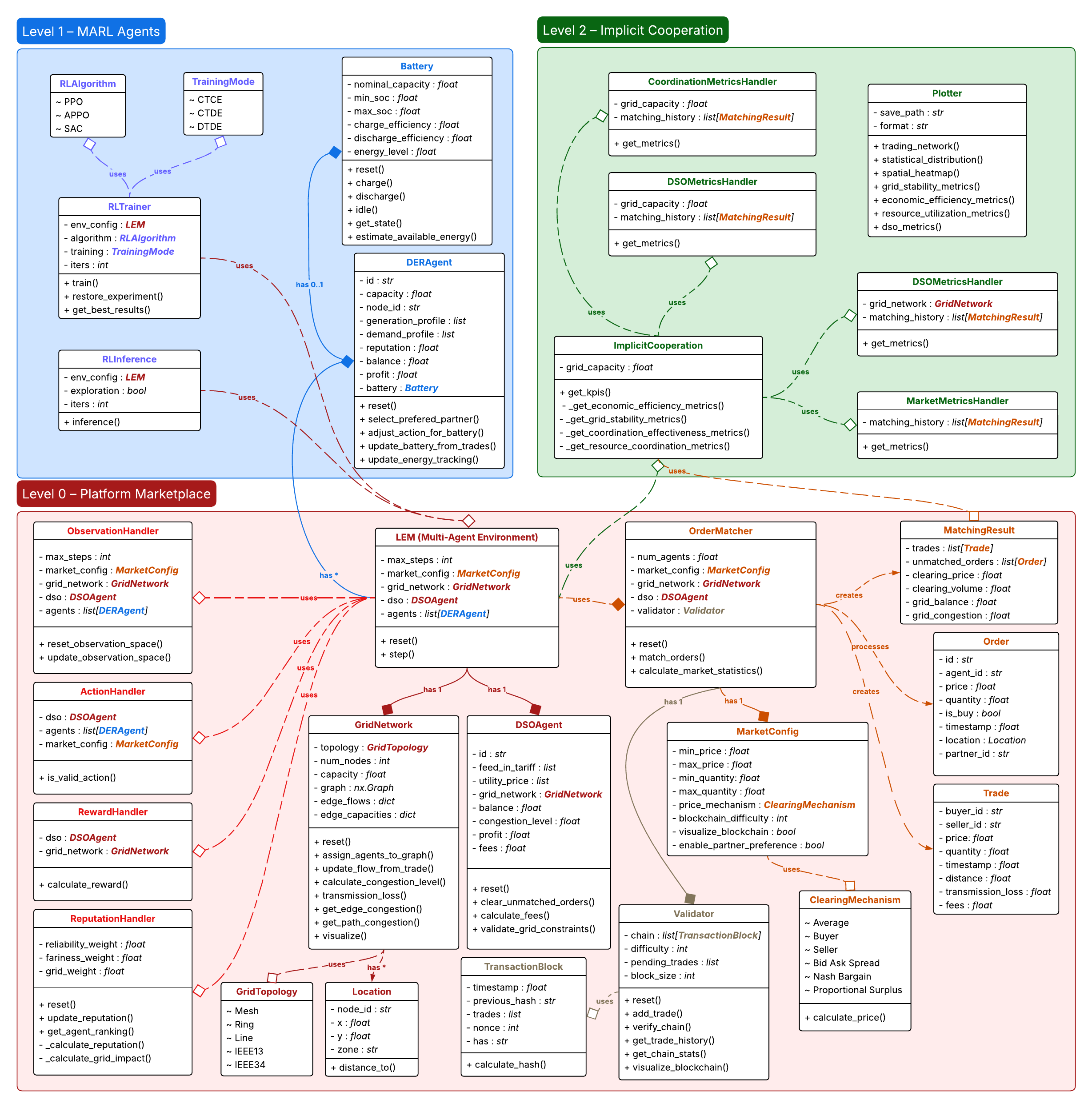}
    \caption{A high-level class diagram illustrating the relationships among the MARLEM framework's core components.}
    \label{fig:uml-diagram}
\end{figure}

The framework is based on a modular, three-level architecture that provides a logical separation between the environment, the actors, and the emergent system dynamics (see Fig. \ref{fig:uml-diagram}). This separation of concerns represents a key design principle, affording high flexibility and extensibility. It permits researchers to isolate and investigate individual components, such as substituting market mechanisms or agent learning algorithms, without necessitating alterations to the remainder of the system. This modularity is fundamental to the framework's role as a versatile research tool, enabling systematic ablation studies and comparative analyses of different LEM configurations.

\begin{itemize}
    \item \textbf{Level 0 -- Platform Marketplace:} This foundational layer simulates the physical and economic environment wherein the agents operate. It is not a passive backdrop but rather an active component possessing its own complex dynamics. This level consists of the \emph{Market Module}, which provides a highly configurable platform for clearing energy trades, and the \emph{Grid Module}, which models the physical power network with veridical constraints. This layer serves to ground the simulation in physical reality by enforcing established physical laws (e.g., congestion and transmission  according to the grid topology) and market regulations, thereby creating a rich, high-fidelity environment that agents must learn to navigate. The fidelity of this layer is essential, as it directly influences the complexity of the policies agents must learn to achieve optimal performance. Detailed technical specifications are provided in Section \ref{1-sec:3.3-lem-design} for the \emph{Market Module} and Section \ref{1-sec:3.5-grid-model} for the \emph{Grid Module}.

    \item \textbf{Level 1 -- MARL Agents:} This layer contains the autonomous agent entities, which constitute the primary actors within the simulation. Each agent is modeled as an independent, economically self-interested participant equipped with its own DERs, such as a battery, and unique generation and demand profiles. The agents learn their policies for market participation based on local observations, with the objective of maximizing their individual rewards. The complexity at this level emerges from the agents' learning processes and their strategic interactions with the market and with one another. The central research challenge addressed by this framework lies in understanding the tension between the agents' pursuit of individual rewards and the achievement of desirable collective outcomes. The complete agent model, including battery dynamics and state management, is detailed Section \ref{1-sec:3.4-agent-model}.

    \item \textbf{Level 2 -- Implicit Cooperation:} This conceptual layer represents the principal scientific focus of the framework. It is not an explicit controller or algorithm but rather the emergent system-level coordination that arises from the engineered interactions between Level 0 and Level 1. This cooperation is guided implicitly through the design of the information structure (i.e., the content of agent observations) and the incentive structure (i.e., the formulation of the reward function). The framework provides the necessary instruments to both foster and quantify this emergent behavior, allowing the study of decentralized coordination. The implementation details are in Section \ref{1-sec:3.6-implicit-cooperation}.
\end{itemize}

\subsection{Formulation of the decentralized partially observable Markov decision process}
\label{1-sec:3.2-decpomdp}

The simulation is orchestrated by the \emph{Environment Module}, which consists primarily of a Gymnasium multi-agent environment that encapsulates the entire LEM. To address the complexities of a decentralized LEM, the problem is formally modeled as a Dec-POMDP. The Dec-POMDP is defined by the tuple $\langle I, \mathcal{S}, A, T, R, \Omega, O \rangle$, where $I$ denotes the set of agents participating in the market, $\mathcal{S}$ the set of environment states, $A$ the joint action space, $T$ the state transition function, $R$ the reward function, $\Omega$ the joint observation space, and $O$ the observation function. Subindex $i$ denotes the state, space, or function for agent $i \in I$.

This formulation is particularly well-suited as it explicitly acknowledges the core challenges of the problem: each agent must make decisions under uncertainty based on its own partial observations within a non-stationary environment where the state transition is contingent upon the joint action of all agents. Furthermore, the formulation provides a robust mathematical lens through which to analyze the LEM:

\begin{itemize}
    \item Partial observability, captured by $\Omega$ and $O(o | s', a)$, directly models the information constraints fundamental to a decentralized system. An agent cannot know the private internal state (e.g., battery level, financial status) of its peers, reflecting both privacy concerns and the physical reality of distributed systems.

    \item Non-stationarity arises from $T(s' | s, a)$, which is conditioned on the joint action $\{a_1 \in A_1, \dots, a_i \in A_i\} \; \forall \;i \in I$. As each agent independently updates its policy, the environment's dynamics appear to shift from the perspective of any single agent.

    \item $R(s, a)$ is designed to tackle the credit assignment problem by providing signals that help agents correlate their individual actions with collective outcomes, a crucial element for fostering implicit cooperation.
\end{itemize}

\subsubsection{Environment state space ($\mathcal{S}$) and transition function ($T$)}

The environment state $s \in \mathcal{S}$ is a comprehensive snapshot of the entire LEM at a given trading period $t \in \{1, \dots, \tau\}$, although it is never fully visible to any single agent. It is composed of the temporal state (e.g., time of day), the complete market state (all orders and trade histories), the physical grid state (congestion on all lines), and the private internal states of all agents.

The state transition function $T(s' | s, a)$ is implicitly defined by the deterministic, sequential logic within the environment's step method. Given a joint action $\{a_1 \in A_1, \dots, a_i \in A_i\} \; \forall \;i \in I$, the environment transitions to a new state ($s' \in \mathcal{S}$) by first processing agent actions into formal orders, then clearing the market, updating all agent and grid states based on the resulting trades, and finally advancing the simulation clock. This procedural definition of the state transition ensures a consistent and reproducible evolution of the environment (see Section \ref{1-sec:3.9-simulation-workflow}).

\subsubsection{Joint observation space ($\Omega$)}

The design of the observation space is of critical importance for enabling implicit cooperation. To facilitate learning in a partially observable environment, each agent $i \in I$ receives a local observation vector $o_i \in \Omega_i$ that constitutes a precisely structured subset of the global state. It provides a sufficiently rich signal for decision-making without violating the principles of decentralization and privacy. The observation vector includes:

\begin{itemize}
    \item \textbf{Market signals (public information):} Publicly available information accessible to any market participant. This encompasses the last clearing price and volume, anonymized statistics regarding P2P versus DSO trading volumes, the prevailing DSO buy (feed-in tariff) and sell (utility) prices, and the reputation score of each agent. These signals provide a common ground of information, reducing uncertainty and allowing agents to form shared expectations about the market's state.  

    \item \textbf{Agent-specific signals (private information):} The agent's private information, which remains unobservable to other agents, thereby preserving privacy. This set of signals includes its current energy generation and demand forecasts, the precise SoC of its battery, its cumulative profit, and its dynamic reputation score. This private information is essential for the agent to tailor its strategy to its own specific circumstances and constraints.

    \item \textbf{Implicit cooperation KPIs (shared signals):} A central and innovative feature of this framework is the inclusion of a selection of system-level KPIs  within the observation vector, such as social welfare, grid congestion, and supply-demand imbalance (see Section \ref{1-sec:3.6-implicit-cooperation}). These KPIs work as a shared public signal that provides all agents with a consistent representation of the overall market health (analogous to a stock market index), thereby allowing them to learn the correlation between their local actions and desirable global outcomes, even without direct communication. For example, an agent can learn that actions contributing to an increase in the observed grid congestion metric may lead to a lower future reward.
\end{itemize}

\subsubsection{Action space ($A$)}

Each agent's action $a_i \in A_i$ is defined as a continuous vector, a formulation that permits fine-grained bidding strategies far more expressive than discrete action spaces. The four components are:  

\begin{equation}
    \label{eq:action-space}
    a_i = \langle p_{bid}, q_{bid}, \alpha, \beta \rangle
\end{equation}

where $p_{bid}$ represents the bid price, $q_{bid}$ denotes the bid quantity, $\alpha \in \{0, 1\}$ indicates the order type (0 for sell, 1 for buy), and $\beta \in \{0, 1, \dots, |I + 1|\}$ is an index representing a preferred trading partner (including the DSO and no preference). The inclusion of $\beta$ as an action component is another novel feature, enabling agents to learn not only what to bid, but also with whom to transact. This facilitates the study of more complex social dynamics beyond simple price-based interactions, such as the formation of trust networks, stable trading coalitions, or community energy groups.

To ensure that actions are valid within the simulation context, two layers of constraints are applied. First, the raw continuous values for price and quantity produced by the agent's policy are clipped to predefined market bounds ($ p_{bid} \in [p_{min}, p_{max}]$, and $q_{bid} \in [q_{min}, q_{max}]$). Second, the agent's requested $q_{bid}$ is then constrained for physical feasibility while considering battery limitations. The framework calculates the maximum energy an agent can physically sell (based on its surplus generation and dischargeable battery capacity) or buy (based on its deficit and chargeable battery capacity).

\subsubsection{Reward function ($R$)}

The reward function $R$ is the primary mechanism through which implicit cooperation is fostered, translating the high-level objective of system stability into a concrete, optimizable signal for each agent. The reward for agent $i \in I$ is a multi-objective function designed to balance individual economic incentives with system-level stability goals:

\begin{equation}
\label{eq:reward-function}
    R_i = R_{base,i} \cdot (1 + f_{coop} \cdot f_{contrib,i}) - \gamma_{DSO,i} - \gamma_{UD,i}
\end{equation}

These components are engineered to operate jointly:

\begin{itemize}
\item $R_{base,i}$: This is a weighted sum representing the agent's individual performance, predominantly influenced by its economic profit from trades. This term drives the agent's self-interested behavior, ensuring it learns effective bidding strategies to operate on its own behalf. $R_{base,i}$ is a composite function that includes the following components:

\begin{itemize}
    \item \textbf{Economic component:} Based on the agent's profit from successful trades. This component incentivizes agents to learn effective bidding strategies.

    \item \textbf{Grid balance component:} Rewards agents for actions that help to reduce the overall grid imbalance (i.e., buying during times of surplus and selling during times of deficit).

    \item \textbf{Resource allocation component:} Rewards efficient use of resources (e.g., full execution of orders, and minimal unused capacity).

    \item \textbf{Trading component:} Rewards the agent's participation in the market during the trading period $t \in \{1, \dots, \tau\}$.

    \item \textbf{Stability component:} Rewards behaviors that contribute to long-term market stability (e.g., price consistency, and grid state improvement).
\end{itemize}

\item $f_{coop}$: This market-wide cooperation factor is calculated from system KPIs and functions as a multiplicative bonus on the agent's base reward. Agents whose actions improve the overall market health (e.g., increase social welfare, reduce congestion) receive a significantly higher reward. Simultaneously, when the market exhibits stable and efficient characteristics (e.g., low congestion, high social welfare), all participating agents receive an augmented reward. This creates a shared incentive for system-level improvements, analogous to a common-pool resource where the market's health is a resource all agents benefit from maintaining. In this way, the emergence of implicit cooperation is encouraged without the need for explicit coordination.

\item $f_{contrib,i}$: This factor refines the cooperation incentive by quantifying the degree to which agent $i$'s specific actions contributed to the positive state of the market. It serves as an effective credit assignment mechanism, helping an individual agent to understand its specific positive or negative impact on the collective. For instance, an agent that sells energy during a grid-wide deficit, thereby contributing to grid stability, will receive a higher contribution score than an agent that sells during a surplus.

\item $\gamma_{DSO,i}$ and $\gamma_{UD,i}$: These are linear penalties applied for trading with the DSO and for the unmet demand, encouraging agents to prioritize P2P transactions and fostering a self-sufficient local market.
\end{itemize}

This multi-faceted reward structure is the primary mechanism through which implicit cooperation is fostered, guiding the decentralized learning processes of individual agents toward emergent, system-level coordination.

\subsection{LEM design and trading mechanisms}
\label{1-sec:3.3-lem-design}

The \emph{Market Module} provides a realistic and flexible simulation of a decentralized LEM. Its primary novelty lies in a hybrid matching mechanism that combines agent-driven partner preferences with a classic price-based auction. This is complemented by configurable operational rules and integrated mechanisms for trust and security, which are critical for the real-world viability of such systems. The market operates in discrete trading periods $t \in \{1, \dots, \tau\}$, following a systematic workflow from order submission to final validation (see Algorithm \ref{alg:market_workflow}).

\begin{algorithm}[h!]
\caption{Market operation workflow at trading period $t \in \{1, \dots, \tau\}$}
\label{alg:market_workflow}
\begin{algorithmic}[1]
    \State \textbf{Input:} Set of all agents $I$, joint action $\{a_1 \in A_1, \dots, a_i \in A_i\} \; \forall \;i \in I$, current market state $s \in \mathcal{S}$.
    \State \textbf{Output:} Set of executed trades, updated market state $s'\in \mathcal{S}$.
    \Statex
    \Statex \Comment{\textit{Phase 1: Order Collection and Validation}}
    \State $\mathcal{O}_t \gets \emptyset$ \Comment{Initialize empty set of valid orders}
    \ForAll{agent $i \in I$}
        \State $o_i \gets \text{ActionToOrder}(a_i)$ \Comment{Convert raw action to a formal order structure}
        \If{$\text{ValidateOrder}(o_i, s)$ is True}
            \State $\mathcal{O}_t \gets \mathcal{O}_t \cup \{o_i\}$
        \EndIf
    \EndFor
    \Statex
    \Statex \Comment{\textit{Phase 2: P2P Market Clearing}}
    \State $trade_{pref}, \mathcal{O}_{rem} \gets \text{PreferenceMatch}(\mathcal{O}_t)$ \Comment{Stage 1: Match mutual preferences}
    \State $trade_{price}, \mathcal{O}_{unm} \gets \text{PriceMatch}(\mathcal{O}_{rem})$ \Comment{Stage 2: Clear via CDA}
    \State
    \State $trade_{p2p} \gets trade_{pref} \cup trade_{price}$
    \ForAll{trade $\in trade_{p2p}$}
        \State $\text{UpdateAgentStates}(trade)$ \Comment{Update energy, profit for buyer and seller}
    \EndFor
    \Statex
    \Statex \Comment{\textit{Phase 3: DSO Clearing and Finalization}}
    \State $trade_{dso} \gets \text{DSOClear}(\mathcal{O}_{unm})$ \Comment{Stage 3: Clear remaining orders with DSO}
    \ForAll{trade $\in trade_{dso}$}
        \State $\text{UpdateAgentStates}(trade)$
    \EndFor
    \State
    \State $trade \gets trade_{p2p} \cup trade_{dso}$ \Comment{Aggregate all trades from the trading period}
    \State $\text{ValidateBlock}(trade)$ \Comment{Simulate blockchain validation for the block of trades}
    \State $s' \gets \text{UpdateMarketState}(trade, s)$ \Comment{Update global prices, volumes, reputations}
    \State $\upzeta_{t+1} \gets \text{UpdateGridState}(trade, s)$ \Comment{Update congestion on the affected edges}
    \State
    \State \textbf{return} $trade, s'$
\end{algorithmic}
\end{algorithm}

\subsubsection{Role of the distribution system operator}
\label{sec:dso}

The DSO is a critical component that fulfills multiple roles essential for market stability and operation. First, it serves as the market maker of last resort, guaranteeing market clearing at every trading period $t \in \{1, \dots, \tau\}$. Second, it acts as a proxy for a real-world grid operator, managing grid stability and collecting fees for grid usage. The DSO maintains a balance ($B_{DSO}$) representing the net energy position (see (\ref{eq:dso-balance})).

\begin{equation}
    \label{eq:dso-balance}
    B_{DSO} = \sum_{trades} (E_{sold} - E_{bought})
\end{equation}

The total fee $F_{total}$ for a given trade is a composite of charges for congestion, usage per unit of distance of the transmission lines (predefined by the DSO), imbalances, voltage support, thermal limits, and cross-zone trading bonus/penalty. The formulation for $F_{total}$ is as follows:

\begin{subequations} \label{eq:dso-fee}
\begin{equation}
    F_{total} = F_{cong} + F_{trans} + F_{imb} + F_{volt} + F_{thermal} + F_{zone}
\tag{\ref{eq:dso-fee}}
\end{equation}

\begin{align}
    F_{cong} &= f_{cong} \cdot (C_{t} - C_{threshold}) \cdot q_{ij} \cdot p_{ij} \\
    F_{imb} &= f_{imb} \cdot |B_{impact}| \cdot q_{ij} \cdot p_{ij}  \\
    F_{volt} &= \min(V_{drop}, V_{threshold}) \cdot q_{ij} \cdot p_{ij} \\
    F_{thermal} &= \min(f_{thermal}, f_{threshold}) \cdot q_{ij} \cdot p_{ij} \\
    F_{zone} &= - f_{zone} \cdot q_{ij} \cdot p_{ij}
\end{align}
\begin{align}
    V_{drop} &= f_{volt} \cdot d_{ij} \\
    f_{thermal} &= \frac{C_{t} - T_{threshold}}{1 - T_{threshold}}
\end{align}
\begin{align}
    B_{impact} =
    \begin{cases}
    \min\left(\frac{q_{ij}}{|B_{grid}|}, 1\right)  & \text{if DSO buying/selling during excess supply/demand} \\
    -\min\left(\frac{q_{ij}}{|B_{grid}|}, 1\right)  & \text{if DSO buying/selling during excess demand/supply} \\
    0 & \text{for balanced grid}
    \end{cases}
\end{align}
\end{subequations}

where $C_t \in [0, 1]$ is the congestion level at trading period $t \in \{1, \dots, \tau\}$, $C_{threshold} \in [0, 1]$ is the congestion intervention threshold  (80\% grid capacity utilization by default), $B_{impact} \in [-1, 1]$ is the balance impact score, $V_{drop}$ is the voltage drop, $V_{threshold} \in [0, 1]$ is the voltage drop threshold (5\% voltage drop by default), $d_{ij}$ is the distance along the network between agents $i$ and $j \in I$, $T_{threshold} \in [0, 1]$ is the thermal threshold for grid instability (20\% grid capacity utilization for thermal limits by default), $B_{grid} \in [0, 1]$ is the balance of the grid at trading period $t \in \{1, \dots, \tau\}$ (a positive balance indicates an excess supply, and a negative balance indicates an excess demand), $f$ are factors for each charge which can be configured based on grid characteristics, and $q_{ij}$ and $p_{ij}$ are the quantity and price for a trade between two agents (including a trade with the DSO).

This fee structure ensures that agents are incentivized to make grid-friendly trading decisions while providing the DSO with revenue to maintain grid infrastructure and stability. Furthermore, it allows for the analysis of various grid tariffing policies and their impact on P2P trading incentives.

\subsubsection{Multi-stage order matching process}

\begin{figure}[h]
    \centering
    \includegraphics[width=0.9\linewidth]{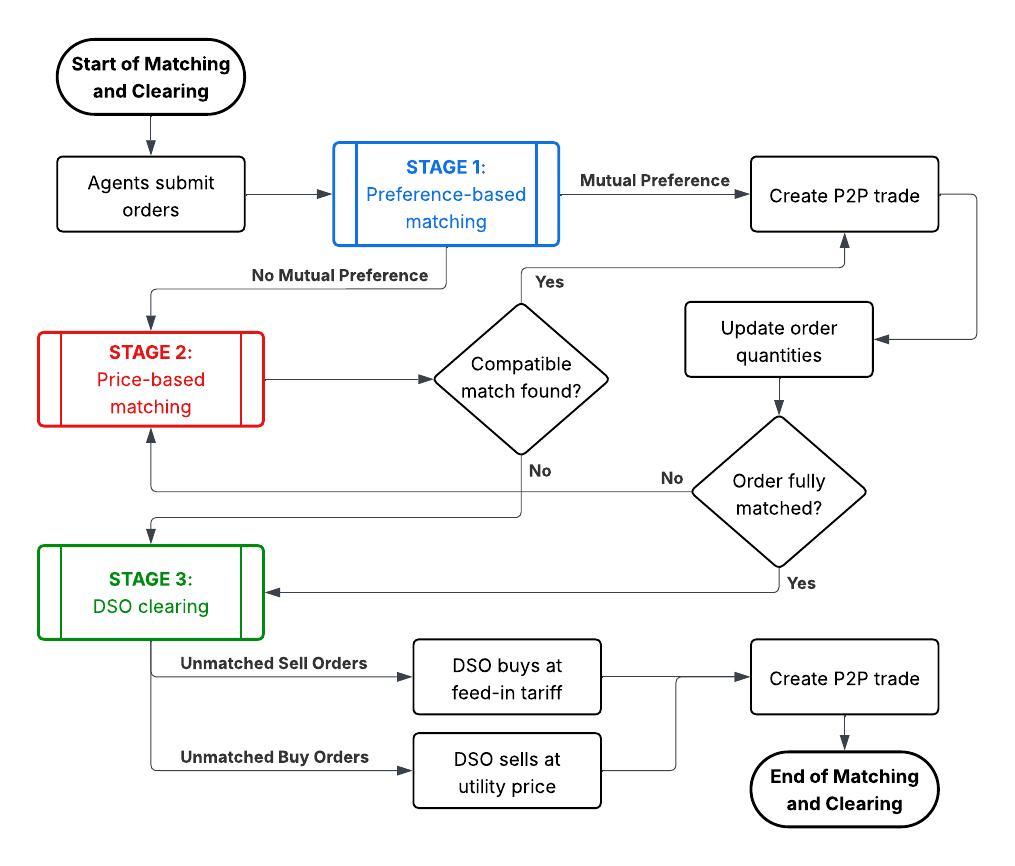}
    \caption{Diagram illustrating the multi-stage order matching and market clearing processes.}
    \label{fig:matching-clearing}
\end{figure}

An innovation of the framework is an algorithm to match the submitted orders, which clears the market utilizing a three-stage, sequential process (see Fig. \ref{fig:matching-clearing}). This hybrid approach is designed to prioritize local and relationship-based trading while concurrently ensuring market liquidity and complete clearing. Moreover, it enables the study of the emergence of cooperative behavior by allowing agents to balance the exploratory nature of an open, anonymous market with the exploitative potential of stable, trusted partnerships. The sequential ordering of the stages is deliberate.

\begin{itemize}
    \item \textbf{Stage 1: Preference-based matching:} The matcher initially attempts to fulfill trades between agents that have mutually specified each other as preferred partners in their respective bid orders. This stage explicitly enables the study of how stable, trust-based trading coalitions may form and whether they enhance or detract from overall market efficiency. By processing these pre-arranged trades first, the system allows established relationships to take priority over the more anonymous and potentially volatile open market, reflecting a form of social efficiency.  

    \item \textbf{Stage 2: Price-based matching:} All remaining orders are processed through a double auction, an established and economically efficient mechanism for anonymous trading. Buy orders are sorted in descending order of price, while sell orders are sorted in ascending order. Trades are subsequently executed for all compatible pairs where the buyer's price meets or exceeds the seller's ($p_{buy} \geq p_{sell}$). Reputation scores are used as a secondary sorting criterion to prioritize more trustworthy agents. 

    \item \textbf{Stage 3: DSO clearing:} Any orders that remain unmatched after the preceding stages are cleared by the DSO. The DSO guarantees that all residual demand is met and all surplus is purchased. However, it executes these transactions at a less favorable prices for the agents: a low feed-in tariff ($p_{fit}$) for buying and a high utility price ($p_{utility}$) for selling. This price disparity, where $p_{fit} < p_{P2P} < p_{utility}$, creates a strong economic incentive for agents to transact within the P2P market in the preceding stages, thereby fostering a self-sufficient local market.
\end{itemize}

\subsubsection{Clearing mechanisms}

A key attribute of the framework is its modularity in market design. The transaction price for a P2P trade is determined by a configurable clearing mechanism, which permits researchers to compare the effects of different pricing rules on agent behavior and market outcomes. Implemented mechanisms include:

\begin{itemize}
    \item \textbf{Average pricing:} A simple and equitable mechanism where the price is set at the midpoint of the bid and ask, ensuring a transparent and computationally efficient settlement (see (\ref{eq:clearing-average})).

    \begin{equation}
        \label{eq:clearing-average}
        p_{ij} = \frac{p_{buy} + p_{sell}}{2}
    \end{equation}

    \item \textbf{Pay-as-bid (seller pricing):} The trade is executed at the seller's bid price ($p_{ij} = p_{seller}$). This rule may incentivize sellers to submit more strategic bids, which can have significant implications for price discovery and overall market efficiency.

    \item \textbf{Pay-as-offer (buyer pricing):} The trade is executed at the buyer's offer price ($p_{ij} = p_{buyer}$), which favors the seller since $p_{buy} \geq p_{sell}$.

    \item \textbf{Nash bargaining pricing:} The price is calculated to equally divide the economic surplus ($p_{buy} - p_{sell}$) between the two agents (see (\ref{eq:clearing-nash})). This approach promotes equitable outcomes and may foster greater long-term market participation by ensuring fairness.

    \begin{equation}
        \label{eq:clearing-nash}
          p_{ij} = p_{sell} + \frac{p_{buy} - p_{sell}}{2}
    \end{equation}

    \item \textbf{Proportional surplus pricing:} The surplus is divided proportionally based on each agent's contribution to the surplus, measured against a reference market price ($p_{ref}$). See (\ref{eq:clearing-proportional}).

    \begin{equation}
        \label{eq:clearing-proportional}
          p_{ij} = p_{sell} + \frac{p_{buy} - p_{ref}}{(p_{buy} - p_{ref}) + (p_{ref} - p_{sell})} \cdot (p_{buy} - p_{sell})
    \end{equation}
\end{itemize}

\subsubsection{Reputation and decentralized validation}

To address challenges of trust and security inherent in decentralized systems, the framework integrates two key features:

\begin{itemize}
    \item \textbf{Reputation system:} The \emph{Reputation Module} assigns a dynamic reputation score $r_i \in [0, 1]$ to each agent $i \in I$, which is a weighted average of their reliability, price fairness, and grid contribution. This score directly influences market outcomes by serving as a tie-breaker in the double auction, rewarding reliable agents with preferential market access. The reputation update process is iterative: after each trading period $t \in \{1, \dots, \tau\}$, the module analyzes completed trades, assesses each agent's grid impact, calculates the new component scores, and updates the agent's overall reputation, incorporating a decay factor to weigh recent behavior more heavily. The reputation score is formally defined as:

    \begin{equation}
    \label{eq:reputation}
        r_i = w_{rel} \cdot \delta_{rel,i} + w_{fair} \cdot \delta_{fair,i} + w_{grid} \cdot \delta_{grid,i}
    \end{equation}
    
    where the scores $\delta$ measure reliability (ratio of matched to total order volume), price fairness (deviation from market clearing price), and grid contribution (whether trades alleviate or exacerbate grid stress) for $t \in \{1, \dots, \tau\}$, and $w$ are the respective weights.  

    \item \textbf{Decentralized validation:} The \emph{Validator Module} emulates a blockchain ledger to ensure transactional integrity without a central arbiter. After each trading period $t \in \{1, \dots, \tau\}$, all executed trades are hashed (SHA-256), aggregated into a block, and validated using a simplified Proof-of-Work consensus mechanism. Once validated, the block is added to an immutable ledger. This procedure creates a transparent record of market activity, a critical feature for ensuring auditability and trust. While Proof-of-Work has known energy inefficiencies, it is employed here as a well-understood and straightforward model for simulating decentralized consensus.
\end{itemize}

\subsection{Agent model and decision-making process}
\label{1-sec:3.4-agent-model}

The agents within the framework are sophisticated entities representing prosumers with their own DERs, each possessing unique characteristics and physical constraints. The \emph{Agent Module} serves as the fundamental building block, encapsulating the agent's internal state, its physical assets, and the logic that governs its interactions with the market.

\subsubsection{Agent architecture and state management}

Each agent is a self-contained entity characterized by a set of core attributes, including a unique identifier, its physical location on the grid, and time-series profiles for its forecasted energy generation $G_t$ and demand $D_t$ for each trading period $t \in \{1, \dots, \tau\}$. The agent's internal state is dynamic, tracking its balance $B_t$ (where a positive balance indicates a surplus of energy, while a negative balance signifies a deficit) and profit, its evolving reputation score $r$, and its cumulative performance in the market. This includes tracking the total demand satisfied versus the demand deferred, providing a measure of the agent's energy security. The agent's balance is estimated in (\ref{eq:der-balance}), where $E_{bought, t}$ and $E_{sold, t}$ are the quantities of energy bought and sold in the market at trading period $t \in \{1, \dots, \tau\}$, respectively.

\begin{equation}
    \label{eq:der-balance}
    B_{t} = (G_t - D_t) + (E_{bought, t} - E_{sold, t})
\end{equation}

A key innovation of our agent model is the integration of a partner selection mechanism. Agents can learn to select preferred trading partners through their RL policy, allowing for the emergence of complex, reputation-based trading relationships. When not controlled by the RL policy, agents can revert to a rule-based strategy that prioritizes partners based on a weighted score of high reputation and short grid distance, promoting both trustworthy and efficient trades.  

The agent's workflow at each trading period $t \in \{1, \dots, \tau\}$ is systematic: it processes the action from its policy, adjusts it for physical constraints, creates a formal market order, and upon market clearing, updates its entire internal state based on the trade outcomes. For example, the cumulative profit is updated as in (\ref{eq:der-profit}).  

\begin{equation}
    \label{eq:der-profit}
    \text{profit}_{t+1} = \text{profit}_t + \sum_{trades}(p_{sell} \cdot q_{sell} - p_{buy} \cdot q_{buy})
\end{equation}

\subsubsection{Physically constrained decision-making}

A central feature of the agent model is the explicit enforcement of physical constraints, which grounds the agent's learned economic policies in physical reality. Before a raw action from the agent's policy is submitted to the market, the agent's action is constrained to ensure its feasibility, preventing an agent from attempting to sell energy it does not possess or purchase energy it cannot store.  
The maximum sellable ($q_{sell,max}$) and purchasable ($q_{buy,max}$) quantities for an agent at trading period $t \in \{1, \dots, \tau\}$ are calculated in (\ref{eq:der-battery-clip}) based on its internal energy balance and its battery state, where $C_{nom}$ is the nominal capacity of the battery and $SoC_{min}$ and $SoC_{max}$ its operational state-of-charge (SoC) limits, and $E_{dischargeable}$ and $E_{chargeable}$ represent the physically available energy and storage capacity in the battery at that moment, respectively.

\begin{subequations}
    \label{eq:der-battery-clip}
    \begin{align}
        q_{sell,max} &= \max(0, G_t - D_t) + E_{dischargeable} \\
        q_{buy,max} &= \max(0, D_t - G_t) + E_{chargeable} \\
        E_{dischargeable} &= (SoC_t - SoC_{min}) \cdot C_{nom} \\
        E_{chargeable} &= (SoC_{max} - SoC_t) \cdot C_{nom}
    \end{align}
\end{subequations}

\subsubsection{Battery storage dynamics}

The battery is an optional component defined by a set of key physical parameters that govern its behavior: $C_{nom}$, $SoC_{min}$ and $SoC_{max}$, and its charging and discharging efficiencies ($\eta_{charge} \in [0, 1]$ and $\eta_{discharge} \in [0, 1]$).

The energy level of the battery ($E_{bat, t} \mid SoC_{min} \cdot C_{nom} \leq E_{bat, t} \leq SoC_{max} \cdot C_{nom}$) evolves at each trading period $t \in \{1, \dots, \tau\}$ based on the net energy flow resulting from the agent's internal balance and market activities.  
The charging and discharging dynamics account for energy losses inherent in the storage process. When charging with an amount of energy $E_{in}$, the stored energy increases. Conversely, when a useful amount of energy $E_{out}$ is delivered by the battery, the stored energy is depleted by a greater amount to account for discharge inefficiency (see (\ref{eq:battery-in-out})).

\begin{align}
    \label{eq:battery-in-out}
    E_{bat, t+1} =
    \begin{cases}
    \min( E_{bat, t} + E_{in} \cdot \eta_{charge}, SoC_{max} \cdot C_{nom}) \\
    \max( E_{bat, t} - \frac{E_{out}}{\eta_{dicharge}}, SoC_{min} \cdot C_{nom})
    \end{cases}
\end{align}

The charge and discharge formulation in (\ref{eq:battery-in-out}) ensure that these operations are always constrained by the battery's instantaneous SoC and its maximum charge/discharge rates. This detailed modeling is essential for accurately capturing the value of energy storage as a source of flexibility in the LEM.

\subsubsection{Time series profiles}

The behavior and decision-making context for each agent are primarily driven by their generation and demand profiles. These time-series data are generated by a dedicated algorithm designed to produce realistic and stochastic inputs. The module can either load data from external files, allowing for experiments based on real-world datasets, or generate it procedurally.

\begin{itemize}
    \item \textbf{Generation profiles:} To simulate renewable energy sources, generation profiles are designed to mimic the output of a solar photovoltaic system. The generation $G_t$ is modeled as a function of the agent's capacity, a base solar irradiance curve (modeled as a Gaussian function centered at midday), and stochastic noise components representing atmospheric effects:

    \begin{equation}
        \label{eq:profile-gen}
        G_t = C_{nom, i} \cdot \lambda_t \cdot (1 + \varepsilon_{noise} + \varepsilon_{cloud})
    \end{equation}

    where $C_{nom, i}$ is the nominal capacity (solar photovoltaic system plus battery) of agent $i \in I$, $\lambda_t$ is the solar irradiance at trading period $t \in \{1, \dots, \tau\}$ modeled as a Gaussian function centered at noon, $\varepsilon_{noise}$ is a random noise term, and $\varepsilon_{cloud}$ is a term that simulates the effect of cloud cover, which can introduce larger, more correlated variations.

    \item \textbf{Demand profiles:} Demand profiles are generated to resemble typical consumption patterns, with peaks in the morning and evening. This creates a natural temporal mismatch with solar generation, thereby establishing the fundamental economic driver for energy trading and storage.
\end{itemize}

This algorithm is also used to replicate the DSO pricing profiles in the \emph{Market Module} (recall Section \ref{sec:dso}). The feed-in tariff and utility price offered by the DSO are generated as step functions that reflect typical wholesale market price variations, with higher prices during peak demand hours. The algorithm ensures that the utility price is always higher than the feed-in tariff, establishing the economic incentive for agents to engage in P2P trading. To enhance realism, all generated profiles are passed through a smoothing filter to reduce abrupt changes and create more natural time-series data.

\subsection{Physical grid model and constraints}
\label{1-sec:3.5-grid-model}

To capture the critical interplay between market activities and physical infrastructure, the framework incorporates a \emph{Grid Module}. This component is indispensable for studying the coupling between economic decisions and their real-world consequences on the stability and efficiency of the electrical grid. It grounds the LEM in physical reality, forcing agents to learn policies that are not only profitable but also physically viable.

\subsubsection{Grid topology and agent allocation}

The physical power distribution network is modeled as a weighted, undirected graph $\mathcal{G} = \langle \mathcal{V}, \xi \rangle$, where agents are located at nodes $v \in \mathcal{V}$ and interconnected by edges $e \in \xi$. A key feature is its support for multiple topology types:

\begin{itemize}
    \item \textbf{Predefined topologies:} For standardized, reproducible experiments, the framework includes canonical models from electrical engineering literature, such as the IEEE 13-node and IEEE 34-node test feeders. 

    \item \textbf{Procedural topologies:} For systematic analysis of network structure, algorithmically generated networks like mesh, ring, and line topologies can be created.
\end{itemize}

Each agent is assigned a location corresponding to a specific node $v \in \mathcal{V}$. The electrical distance between two agents $d_{ij}$ is then calculated not as a simple Euclidean distance but as the length of the shortest path through the network graph between their respective nodes ($v_i$ and $v_j$). This distance is a critical input for calculating the physical effects of energy transfers.

\subsubsection{Modeling physical constraints}

The \emph{Grid Module} imposes two fundamental physical constraints on all energy transactions, which have direct economic consequences and create complex learning challenges for the agents.

\begin{itemize}
    \item \textbf{Transmission losses:} Energy is inevitably dissipated during transport over power lines. The framework models these losses ($\ell_{ij}$) for a trade of quantity $q_{ij}$ over an electrical distance $d_{ij}$ as a linear function considering a configurable loss factor $\kappa$ (see (\ref{eq:grid-losses})).

    \begin{equation}
        \label{eq:grid-losses}
        \ell_{ij} = d_{ij} \cdot q_{ij} \cdot \kappa
    \end{equation}
    
    These losses $\ell_{ij}$ derates $q_{ij}$ as it is transported through the grid, and are incorporated into trading fees by the DSO, creating a direct economic incentive for agents to prioritize local transactions with nearby peers. This is a primary driver for the formation of geographically clustered, community-level energy markets.

    \item \textbf{Congestion:} The power flow $\mathcal{F}$ on each power line $e \in \xi$ is monitored relative to its maximum capacity. The total grid capacity is distributed among the edges ($C_e$), with lower-impedance (shorter) edges typically having higher capacity. The congestion level $\upzeta$ at an edge $e \in \xi$ is the ratio between the current power flow at that edge and its maximum capacity (see (\ref{eq:grid-congestion})).

    \begin{equation}
        \label{eq:grid-congestion}
        \upzeta_e = \frac{\mathcal{F}_e}{\max(C_e)}
    \end{equation}
    
    After each trading period $t \in \{1, \dots, \tau\}$, the power flow $\mathcal{F}$ on each power line $e \in \xi$ along the shortest path of every executed trade is updated. Crucially, the resulting grid-wide congestion level $\upzeta$ is then fed back into the agents' observation space as a system-level KPI. This mechanism allows agents to learn strategies that collectively alleviate grid stress, such as reducing power transfers along highly loaded corridors. This demonstrates a form of emergent, decentralized congestion management, a critical function for the stability of future grid operations.
\end{itemize}

\subsubsection{Dynamic grid state updates and management}

The state of the grid is not static but evolves dynamically in response to market activities. This is managed through a systematic update process that occurs after each trading period $t \in \{1, \dots, \tau\}$.

\begin{itemize}
    \item \textbf{Power flow update process:} For each trade, the algorithm first identifies the shortest path between the buyer and seller on the network graph. Then, the power flow $\mathcal{F}$ on every edge $e \in \xi$ along that path is incremented by the trade quantity $q_{ij}$ (considering transmission losses). This ensures that the physical impact of every economic transaction is reflected in the grid's state. Trades that would violate capacity constraints can be penalized or disallowed by the DSO.

    \item \textbf{Grid balance calculation:} The overall health of the grid is monitored through a grid balance metric $B_{grid}$, which quantifies the net energy surplus or deficit within the LEM. This is calculated in (\ref{eq:grid-balance}) as the sum of all agents' net energy positions. A value of $B_{grid}$ close to zero indicates a self-sufficient market, while large deviations signify a heavy reliance on the external grid, a key indicator of system performance.

    \begin{equation}
        \label{eq:grid-balance}
        B_{grid} = \sum_{i \in I}(G_i - D_i + E_{bought,i} - E_{sold,i})
    \end{equation}
\end{itemize}

\subsubsection{Zonal grid structure}

To enable the study of more complex, large-scale power systems and sophisticated market designs, the framework supports a zonal grid structure. This feature allows the main grid graph $\mathcal{G}$ to be partitioned into a set of distinct, connected subgraphs, known as zones, where each zone $\mathcal{Z} = \langle \mathcal{V}, \xi \rangle \subseteq \mathcal{G}$. Although it is configurable, by default the framework applies a bonus (negative fees) if the buyer and seller are in the same zone.

This zonal definition serves two primary research purposes. First, it allows for the modeling of geographically or electrically distinct areas within a larger grid, each with potentially different characteristics. Second, it facilitates the implementation of advanced market mechanisms, such as location-based marginal pricing or zonal congestion management schemes. For example, trades occurring between agents in different zones can be subjected to additional fees or constraints, simulating the costs and limitations of transferring power across wider areas. This provides a comprehensive tool for analyzing the scalability and hierarchical control of decentralized energy markets.

\subsection{Implicit cooperation and performance measurement}
\label{1-sec:3.6-implicit-cooperation}

A central hypothesis of this research is that self-interested agents, operating in a shared environment with well-designed incentives, can learn to cooperate implicitly without direct communication. The \emph{Implicit Cooperation Module} is fundamental to this process. It does not enforce cooperation, but rather measures and incentivizes it by providing agents with system-level information that promotes collective benefits.

The \emph{Implicit Cooperation Module} calculates a suite of KPIs that serve as feedback signals, guiding the agents' learning process toward mutually beneficial outcomes. These KPIs are integrated both into the agents' observation space, providing a shared signal of the system's state, and into the reward function, directly rewarding actions that contribute to market health. This creates a feedback loop where individual actions are shaped by their impact on the collective.

\subsubsection{The coordination feedback loop}

The mechanism for fostering implicit cooperation operates through a closed feedback loop that integrates the KPIs into the learning process, functioning as the shared signals at each trading period. By including these KPIs in each agent's observation vector, all agents receive a consistent, shared signal about the state of the system.

At the end of a trading period $t \in \{1, \dots, \tau\}$, the \emph{Grid Module} calculates the physical outcomes (e.g., congestion), and the \emph{Market Module} records the economic outcomes (e.g., prices, volumes). This data is then consumed by the \emph{Implicit Cooperation Module}, which computes the system-level KPIs. These KPIs are not merely logged for post-hoc analysis but are actively injected back into the environment in two ways:

\begin{enumerate}
    \item \textbf{As shared observations:} A curated vector of KPIs is appended to each agent's local observation. This provides all agents with a consistent, shared signal about the collective state of the system, allowing them to learn the correlation between their local actions and global outcomes. 

    \item \textbf{As reward components:} The KPIs are used as direct inputs to the calculation of the $f_{coop}$ and $f_{contrib, i}$ terms in the reward function (recall (\ref{eq:reward-function})). This creates a direct and tangible incentive for agents to take actions that improve these system-level metrics, effectively aligning individual self-interest with the collective good.
\end{enumerate}

This continuous flow of information from global state to local signals is what enables agents to learn coordinated, pro-social behaviors without requiring explicit communication or a central controller. The feedback mechanism adapts to changing market conditions.

\subsubsection{Key performance indicators}

The framework calculates a comprehensive suite of KPIs, categorized to provide a multi-faceted view of system performance. The following are the indicators used to measure emergent cooperation:

\begin{enumerate}
    \item \textbf{Economic efficiency KPIs:} This set of metrics measures the market's ability to create value and facilitate efficient price discovery.
    
    \begin{itemize}
        \item \textbf{Social welfare:} The total economic value of all trades, representing the sum of consumer and producer surplus. It serves as the primary indicator of overall market efficiency.

        \begin{equation}
            \label{eq:social-welfare}
            \text{Social Welfare} = \sum_{trades} p_{trade} \cdot q_{trade}
        \end{equation}

        \item \textbf{Market liquidity:} The total volume of energy traded, indicating market activity and depth.

        \begin{equation}
            \label{eq:market-liquidity}
            \text{Liquidity} = \sum_{trades} q_{trade}
        \end{equation}

        \item \textbf{Average bid-ask spread:} The average difference between buy and sell order prices, measuring market efficiency.

        \begin{equation}
            \label{eq:bid-ask-spread}
            \text{Spread} = \mathbb{E}[p_{ask}] - \mathbb{E}[p_{bid}]
        \end{equation}

        \item \textbf{Price volatility:} The standard deviation of the clearing price over a time window, indicating market stability.
    \end{itemize}

    \item \textbf{Grid stability KPIs:} This set of metrics assesses the physical health and operational efficiency of the electrical grid.
    
    \begin{itemize}
        \item \textbf{Supply-demand imbalance:} The net energy imbalance normalized by grid capacity, measuring how well supply and demand are balanced. A value near zero suggests stable operation that does not strain the wider grid.

        \begin{equation}
            \label{eq:imbalance}
            \text{Imbalance} = \frac{| \sum q_{buy} - \sum q_{sell} |}{C_{grid}}
        \end{equation}

        \item \textbf{Grid gongestion:} The average congestion level across all power lines $e \in \xi$ as defined in (\ref{eq:grid-congestion}), indicating physical stress on the infrastructure. Low congestion levels are indicative of a system operating well within its physical limits, enhancing reliability.

        \item \textbf{Grid balance:} The overall energy balance of the grid, calculated as the difference between total generation and consumption as in (\ref{eq:grid-balance}).
    \end{itemize}

    \item \textbf{Resource coordination KPIs:} These metrics evaluate how effectively DERs are utilized and coordinated.
    
    \begin{itemize}
        \item \textbf{DER self-consumption:} The proportion of total energy transacted that occurs in P2P trades as opposed to with the DSO. A high value is indicative of a self-sufficient and effective local market, reducing reliance on centralized utilities.

        \begin{equation}
            \label{eq:self-consumption}
            \text{Self-Consumption} = \frac{\sum q_{P2P}}{\sum (q_{P2P} + q_{DSO})}
        \end{equation}

        \item \textbf{Flexibility utilization:} The proportion of available flexible energy that is actively utilized in P2P trading. This metric measures how effectively agents utilize their energy flexibility resources (generation surplus, demand deficit, and battery capacity).

        \begin{equation}
            \label{eq:flexibility}
            \text{Flexibility Utilization} = \frac{\sum q_{p2p}}{\sum q_{available}}
        \end{equation}

        where $\sum q_{available}$ is the total available flexibility across all agents, considering the sellable flexibility (surplus generation plus battery discharge capacity), and the buyable flexibility (deficit demand plus battery charge capacity) at a given trading period $t \in \{1, \dots, \tau\}$.
    \end{itemize}

    \item \textbf{Coordination effectiveness KPIs:} These metrics measure the emergence and effectiveness of coordination among agents.
    
    \begin{itemize}
        \item \textbf{Coordination score:} A measure of coordination, reflecting the market's balance. A score approaching 1 indicates perfect system balance.

        \begin{equation}
            \label{eq:coordination}
            \text{Coordination Score} = 1 - \text{Imbalance}
        \end{equation}

        \item \textbf{Coordination convergence:} Measures the stability of trading volumes over a recent window, indicating if a stable, coordinated pattern has emerged. It is calculated similarly to the price volatility metric.
    \end{itemize}
\end{enumerate}

\subsection{Multi-agent reinforcement learning framework}
\label{1-sec:3.7-marl-framework}

While the preceding sections have detailed the environment's design, this subsection describes the MARL framework responsible for training the agents' policies. A significant contribution of our work is the framework's native support for comparing different MARL training and inference paradigms, which is essential for studying the trade-offs between centralization and decentralization in learning and control. This is a core research instrument that allows for the rigorous, side-by-side evaluation of coordination strategies under varying degrees of information sharing and architectural constraints. The entire training and inference pipeline is managed by a dedicated \emph{Trainer Module} and \emph{Inference Module}, respectively, built upon established libraries like Ray RLlib \cite{liang2018}, to ensure robustness, scalability, and reproducibility.

A detailed treatment of the MARL framework, including its full mathematical and implementation specifics, is reserved for a forthcoming paper; the following serves as a comprehensive overview of its key components and capabilities.

\subsubsection{Training and execution paradigms}
\label{1-sec:3.7.1-training-paradigms}

A primary research objective of this work is to evaluate the performance and scalability of different levels of decentralization. To this end, the framework is designed to seamlessly implement and compare three canonical MARL paradigms, each offering a unique balance of performance, scalability, and real-world deployability.

\begin{itemize}
    \item \textbf{Centralized Training, Centralized Execution (CTCE):} In this paradigm, a single, centralized policy has access to the full global state  (concatenated observations of all agents) and outputs a joint action  for all agents simultaneously. This model serves as a theoretical upper bound for system performance, as it represents the ideal of perfect information and fully coordinated execution. While often intractable in real-world systems due to its violation of decentralization principles and exponential scaling challenges, it provides a benchmark for quantifying the performance gap between the theoretical optimum and more practical, decentralized approaches.

    \item \textbf{Centralized Training, Decentralized Execution (CTDE):} This is a popular and practical paradigm that represents a hybrid approach. Agent policies are trained offline in a centralized simulator with access to global information (e.g., the states and actions of other agents, centralized critics). This allows the learning algorithm to solve the credit assignment problem and learn complex, coordinated strategies. During inference, however, each agent's policy operates in a fully decentralized manner, relying only on its own local observation. This approach is highly compelling because it allows for sophisticated, coordinated strategies to be learned offline while still producing policies that are deployable in a realistic, decentralized setting without requiring a central controller at runtime.

    \item \textbf{Decentralized Training, Decentralized Execution (DTDE):} This represents the fully decentralized ideal, most closely mirroring the constraints of many real-world systems. Each agent learns its policy independently, based only on its own stream of local observations and rewards, without a centralized training coordinator. This paradigm is highly scalable and robust to single points of failure. However, it faces significant theoretical challenges related to the non-stationarity of the environment (as other agents' policies are constantly changing) and the difficulty of credit assignment. Our framework's focus on implicit cooperation through shared KPI signals in the observation and reward functions is specifically designed to address these challenges and make DTDE a more viable and effective approach.
\end{itemize}

\subsubsection{Reinforcement learning algorithms and infrastructure}

The framework integrates with high-performance RL libraries to implement the learning algorithms. While the modular design is algorithm-agnostic, our experiments primarily leverage state-of-the-art policy gradient and actor-critic methods suitable for multi-agent environments and continuous action spaces. We highlight three implemented algorithms:

\begin{itemize}
    \item \textbf{Proximal policy optimization (PPO):} A robust on-policy algorithm recognized for its stability and sample efficiency \cite{schulman2017}. Its core innovation is a clipped objective function that prevents destructively large policy updates, making it a strong and reliable baseline for complex control tasks.
    
    \item \textbf{Asynchronous proximal policy optimization (APPO):} APPO enhances scalability for distributed training through off-policy corrections (V-trace), which allows the algorithm to learn from off-policy data more effectively \cite{schulman2017}. This is particularly valuable in multi-agent environments where agents may have different learning rates and exploration strategies, leading to off-policy data that standard PPO cannot handle efficiently.

    \item \textbf{Soft actor-critic (SAC):} An off-policy algorithm that encourages exploration by adding an entropy bonus to the objective \cite{haarnoja2018}. This principled approach to balancing exploration and exploitation makes it highly effective in continuous control tasks, often leading to more robust and performant final policies, which is particularly valuable for discovering novel coordination strategies in the LEM.
\end{itemize}

To manage the complexity of large-scale MARL experiments, the training infrastructure incorporates advanced techniques such as Population-Based Training \cite{jaderberg2017}, which is a hybrid approach that jointly optimizes the neural network weights and hyperparameters of a population of agents, dynamically allocating computational resources to the most promising training runs. This automates the difficult process of hyperparameter tuning and often leads to more robust and higher-performing final policies, which is particularly beneficial in the non-stationary context of MARL. The entire experimental workflow, from training and checkpointing to metrics collection and final inference, is managed to ensure reproducibility and rigorous scientific comparison.

\subsection{Analytics and visualizations}
\label{1-sec:3.8-analytics-visualization}

An \emph{Analytics Module} and a \emph{Visualization Module} are integrated into the framework, providing a suite of tools for processing, analyzing, and plotting the high-dimensional data generated during simulation runs. This module is essential for moving beyond raw numerical outputs to a qualitative understanding of the emergent dynamics within the LEM.

The visualization capabilities are categorized to provide distinct perspectives on the simulation outcomes:

\begin{itemize}
    \item \textbf{Analysis of social and relational dynamics:} To investigate the emergence of trading relationships and market structure, the framework can generate a trading network figure. In this plot, agents are rendered as nodes in a graph, and the executed trades are represented as edges connecting them. The thickness of an edge is proportional to the total volume of energy traded between the two agents, and the grayscale tone of each edge represents the distance between the agents, providing a visualization of the market's social fabric. This allows for the direct observation of how stable trading coalitions and preferred partnerships evolve over time.

    \item \textbf{Economic performance analysis:} To understand the statistical properties of market outcomes, the \emph{Visualization Module} can generate price and volume distribution plots. These histograms and kernel density estimates reveal the distribution of market clearing prices and trade volumes, offering insights into price stability, market liquidity, and the overall efficiency of price discovery under different market mechanisms.

    \item \textbf{Spatio-temporal pattern recognition:} To analyze the physical dimension of market activity, a spatial distribution of trading activity plot can be generated. This visualization renders the grid topology as a heatmap, where the color intensity of different nodes or areas corresponds to the volume of trading activity. This is particularly useful for identifying the emergence of localized trading hubs and for correlating economic activity with physical grid phenomena, such as congestion.

    \item \textbf{Performance dashboards:} For a high-level, comparative overview of different experimental scenarios, the framework provides a suite of metrics dashboards. These composite plots summarize the most critical KPIs from the \emph{Analytics Module} into a single, easily digestible format. Dedicated dashboards are available for grid stability, economic efficiency, resource utilization, and DSO performance, allowing researchers to quickly assess and compare the multi-faceted outcomes of their experiments.
\end{itemize}

\subsection{Complete simulation workflow}
\label{1-sec:3.9-simulation-workflow}

\begin{figure}[h!]
    \centering
    \includegraphics[width=\textwidth, angle=0]{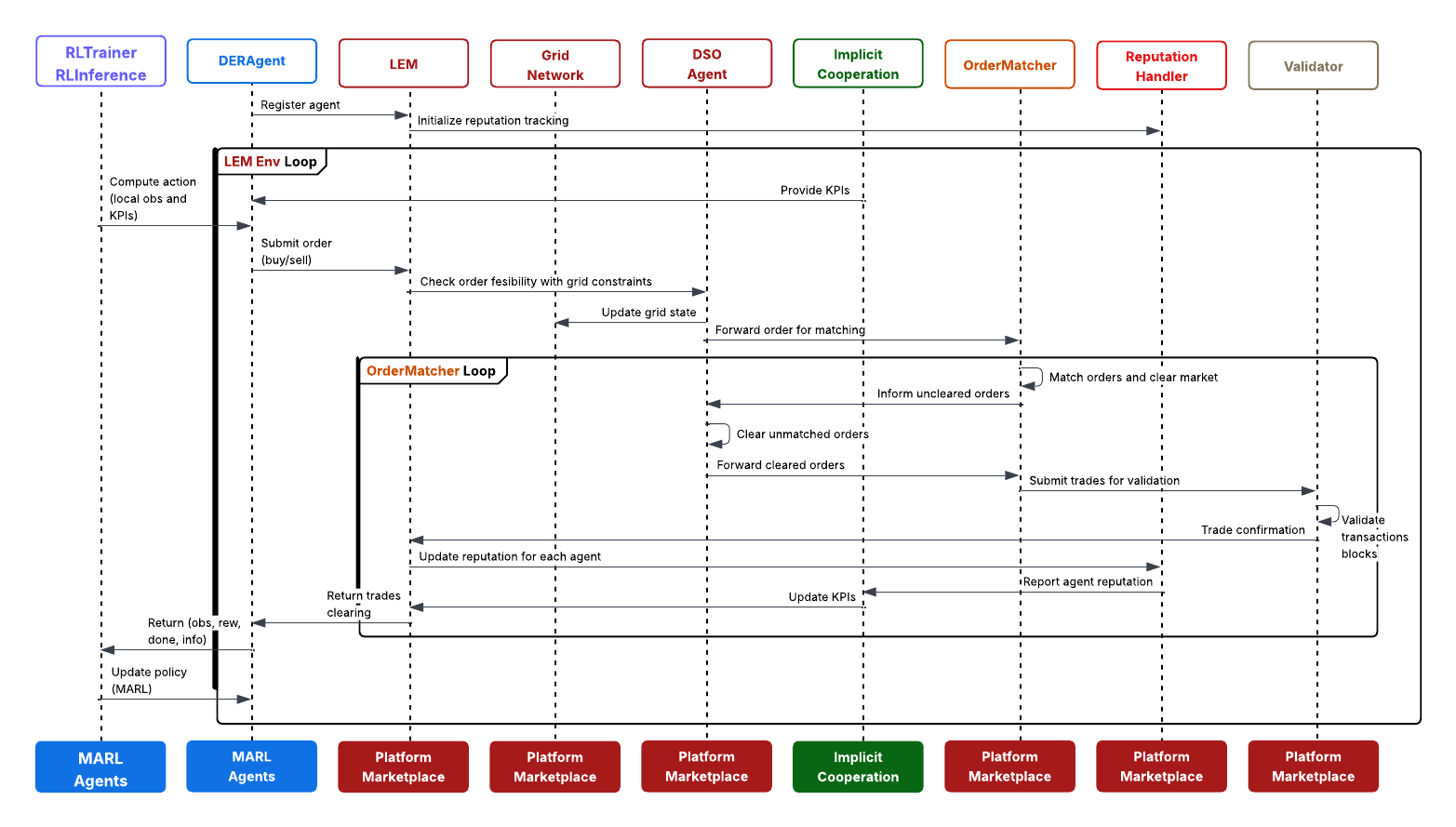}
    \caption{A sequence diagram for the simulation workflow of the MARLEM framework.}
    \label{fig:sequence-diagram}
\end{figure}

The preceding sections have detailed the individual components of the framework; this final subsection elucidates how these modules are orchestrated into a complete, end-to-end simulation loop. The step method within the LEM environment manages this sequence of operations, ensuring that data flows logically between the agents, market, grid, and coordination modules at each discrete trading period $t \in \{1, \dots, \tau\}$. Understanding this workflow is essential for appreciating how the framework maintains state consistency and facilitates the agent learning process. The entire process at each trading period $t \in \{1, \dots, \tau\}$ is conceptualized as a sequence of eight distinct phases, from initial agent decision to the final observation and reward feedback (see Fig. \ref{fig:sequence-diagram}).

A detailed breakdown of the simulation process is as follows:

\begin{enumerate}
    \item \textbf{Action processing and validation:} The workflow begins with the raw action  received from the MARL policies. For each agent, the first and most critical step is to ground its intended action in physical reality. Agents limit each action based on the current generation, demand, and the agent's battery state. Subsequently, the \emph{Action Module} performs a rule-based validation, ensuring the action's components (price, quantity, buy/sell, partner) fall within the predefined market bounds. This two-step validation ensures that only physically and economically plausible actions proceed to the next phase.

    \item \textbf{Order creation and market submission:} Validated actions are then formatted into formal orders. If the agent's policy is not directly controlling partner choice, a rule-based selection method can be invoked here, scoring potential partners based on reputation and grid distance. This phase culminates in a collection of structured orders ready for market submission.

    \item \textbf{Market clearing and trade execution:} The collection of orders is passed to the \emph{Market Module} which executes its three-stage clearing process: it first attempts to match mutually preferred partners, then processes the remaining orders through the price-based double auction (factoring in reputation scores), and finally clears any residual unmatched orders with the DSO. The output of this phase is a list of all executed trades and summary statistics.

    \item \textbf{Energy tracking and battery updates:} The list of all executed trades is used to update the physical state of each agent (e.g., internal energy tracking variables and change in the battery's SoC), and grid state (e.g., balance and congestion).

    \item \textbf{Reward calculation and KPI updates:} With the economic and physical outcomes of the trading period $t \in \{1, \dots, \tau\}$ now determined, the \emph{Implicit Cooperation Module} gathers all necessary data (trades, agent states, grid conditions) to compute the full suite of system-level KPIs. These KPIs are then passed to the reward function, which calculates the final, composite reward for each agent, combining their individual performance with the cooperation bonuses derived from the system-wide KPIs.

    \item \textbf{State transitions and environment updates:} This phase advances the simulation's global state, and the environment's internal clock is incremented. Crucially, agent-level states that depend on the full market outcome, such as financial balances and reputation scores, are updated based on the results from the \emph{Reputation Module}.

    \item \textbf{Observation update and information sharing:} The \emph{Observation Module} constructs the next local observation for each agent ($o_{i, t+1} \in O_i$). This new observation vector is assembled from the newly updated environment state, including the agent's private information, public market signals, and the system-level KPIs calculated in Phase 5. This step closes the information loop, providing agents with the feedback necessary for their next decision.

    \item \textbf{Episode management and termination:} Finally, the environment checks for termination conditions (e.g., reaching the maximum number of trading periods $\tau$ for the episode). If the episode has not concluded, the environment returns the newly generated observations and computed rewards to the MARL algorithm, and the entire workflow repeats for trading period $t + 1 \in \{1, \dots, \tau\}$.
\end{enumerate}

This orchestrated workflow ensures that every agent's action influences the market and grid, the collective outcome is measured and translated into shared signals (KPIs), and those signals are then used to inform and incentivize the agents' decisions in the subsequent trading period. This constitutes the core feedback loop that enables the study of emergent, implicit cooperation within a complex, decentralized system.

\section{Experimental setup}
\label{1-sec:4-experimental-setup}

To evaluate the capabilities of the proposed LEM simulation framework and demonstrate its utility, a computational experiment was designed. This section details the experimental scenario, implementation specifics, and baseline models for comparison. While the framework supports a wide range of experiments, this paper focuses on a single, comprehensive case study: battery storage coordination. This case study is selected for its ability to demonstrate the interplay between individual agent capabilities (energy storage), market dynamics, physical constraints, and the emergence of implicit coordination, thereby utilizing all core modules of the framework. Readers interested in the full suite of implemented case studies (see Section \ref{1-sec:4.4-other-studies}) or wishing to replicate the experiments described here can find detailed configurations and corresponding implementations in the project's dedicated GitHub repository (\href{https://github.com/salazarna/marlem}{https://github.com/salazarna/marlem}).

\subsection{Case study: battery storage coordination}
\label{1-sec:4.1-case-study}

This case study showcase and analyzes the role of battery energy storage systems in decentralized LEMs, leveraging the framework's integrated models.

The primary goal is to analyze how different battery configurations and deployment strategies affect the emergence of implicit coordination among agents (MARL or otherwise), influence overall market efficiency (e.g., price volatility, social welfare), and contribute to grid stability through enhanced flexibility.

The simulation environment is configured with 7 agents operating on a 24-hour daily cycle in a grid network with a mesh topology and a nominal capacity of 1200 kW. A bid-ask spread clearing mechanism is employed, chosen specifically because the inherent difference between buy and sell prices creates clear economic incentives for storage arbitrage, thus highlighting the value that agents learn (or exploit, if they do not learn) from their batteries. The market allows for a wide price range (i.e., 35-280 \$/MWh) and quantity range (i.e., 0.1-200 kWh) to accommodate storage-driven strategies. 

In this paper we showcase two distinct scenarios (although readers can find six comprehensive scenarios in the dedicated GitHub repository), each featuring agents with the same base generation capacity (i.e., 60 kW, with some random variation) and associated profiles but with differences in the deployment of their batteries:

\begin{enumerate}
    \item \textbf{No battery:} A baseline scenario establishing performance without any storage flexibility, forcing agents into real-time market participation only. All agents are configured to have no battery.

    \item \textbf{Strategic battery:} Explores deployment where agent profiles (generation/demand timing) and storage capabilities (varying ratios 0.5-1.2, high 95\% efficiency) are specifically designed a priori to facilitate coordinated load balancing and time-shifting across the community (e.g., agents with morning generation peaks paired with storage potentially serving agents with evening demand peaks).
\end{enumerate}

\subsection{Baseline models}
\label{1-sec:4.2-baseline-models}

To evaluate the performance achieved by the agents within case study, the following baselines are essential for establishing context:

\begin{itemize}
    \item \textbf{Zero-intelligence agents:} This baseline serves as a lower bound on performance. Agents operate without any learning or strategic foresight. At each time step, they submit random bids using a uniform random distribution within price bounds that are valid within the market's price and quantity limits and respect their individual physical constraints (e.g., battery SoC, generation availability). Comparing MARL performance (when trained) against this baseline demonstrates that learned policies are non-trivial and achieve meaningful economic and coordination outcomes beyond random chance. For the purposes of clarity and demonstration of the capabilities of the framework, the results presented in Section \ref{1-sec:5-results-and-discussion} will utilize these zero-intelligence agents to illustrate the impact of different configurations (e.g., battery scenarios).

    \item \textbf{MARL with CTCE paradigm:} As detailed in Section \ref{1-sec:3.7.1-training-paradigms}, this paradigm acts as a theoretical upper bound within the context of the MARL algorithms used. A single policy, trained with access to the full global state and controlling all agents simultaneously using the same underlying MARL algorithm (e.g., PPO, APPO or SAC), represents the best achievable performance under conditions of perfect information and perfect coordination attainable by that algorithm. Comparing decentralized execution results (CTDE or DTDE) against the CTCE benchmark allows for quantifying the performance degradation attributable to partial observability and independent decision making inherent in decentralized execution. While full MARL training is deferred to future work (see Section \ref{1-sec:6-conclusion}), this benchmark remains relevant for contextualizing potential performance.
\end{itemize}

\subsection{Other implemented case studies}
\label{1-sec:4.4-other-studies}

Beyond the focus on battery storage coordination, the framework's versatility is further demonstrated by other fully implemented case studies available in the dedicated GitHub repository. These include:

\begin{itemize}
    \item \textbf{Market mechanism comparison:} Compares six different pricing/clearing rules to assess their impact on efficiency and agent strategies.

    \item \textbf{Agent heterogeneity and market power:} Explores market dynamics under varying levels of agent size concentration (balanced, monopoly, oligopoly).

    \item \textbf{DSO intervention strategies:} Evaluates how different DSO regulatory stances (permissive, moderate, strict) affect market participation and grid stability.

    \item \textbf{Grid topology and congestion effects:} Examines the influence of network structure and capacity limits on trading patterns and coordination.
\end{itemize}

These additional studies, while not detailed here, underscore the framework's capacity as a comprehensive tool for LEM research.

\section{Results and discussion}
\label{1-sec:5-results-and-discussion}

This section presents the analytical capabilities and techno-economic sensitivity of the proposed simulation framework. As outlined in Section \ref{1-sec:4-experimental-setup}, the findings presented here are generated using the zero-intelligence agent baseline. By comparing the \emph{no battery} and \emph{strategic battery} scenarios using non-learning agents that operate under randomized but valid heuristics, we can demonstrate the framework's high-fidelity modeling. This allows us to observe how the physics of the system (the market rules, grid constraints, and agent asset configurations) shape the outcomes, independent of the learned MARL policies. The results from full MARL training, which constitutes the next work of this research, are detailed in Section \ref{1-sec:6-conclusion}.

\subsection{Framework capability demonstration}
\label{1-sec:5.1-framework-demostration}

The core capability of the framework is its ability to quantify the systemic impact of different physical asset configurations. By simulating and contrasting these two scenarios, we can observe how the presence and strategic deployment of storage fundamentally alter market outcomes, even when agent behavior is non-strategic.

\subsubsection{Price volatility and market stability}

\begin{figure}[h]
    \centering
    \begin{subfigure}{\textwidth}
       \centering
       \includegraphics[width=\linewidth, trim=0cm 0cm 0cm 1cm, clip]{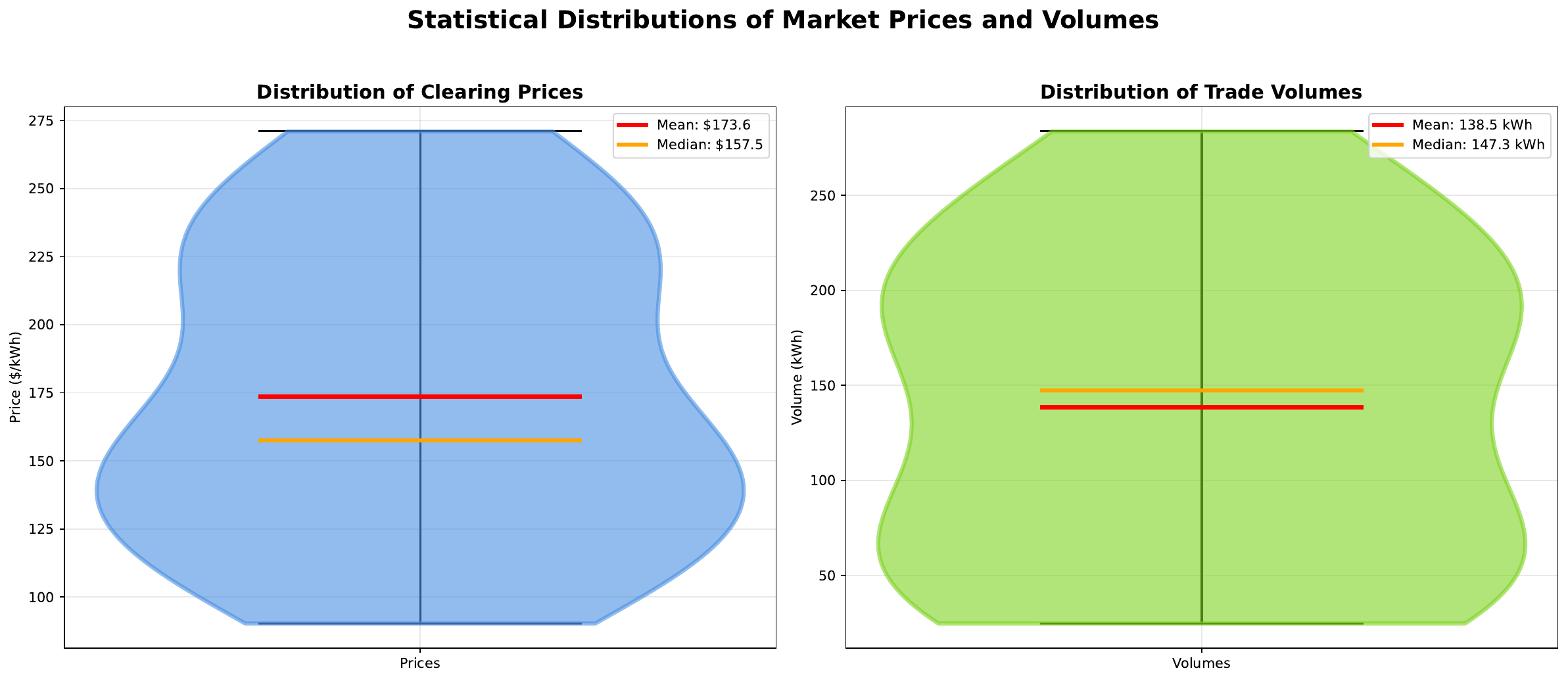}
       \caption{No battery scenario.}
       \label{fig:statistical-distribution-no-battery} 
    \end{subfigure}
    \begin{subfigure}{\textwidth}
       \centering
       \includegraphics[width=\linewidth, trim=0cm 0cm 0cm 1cm, clip]{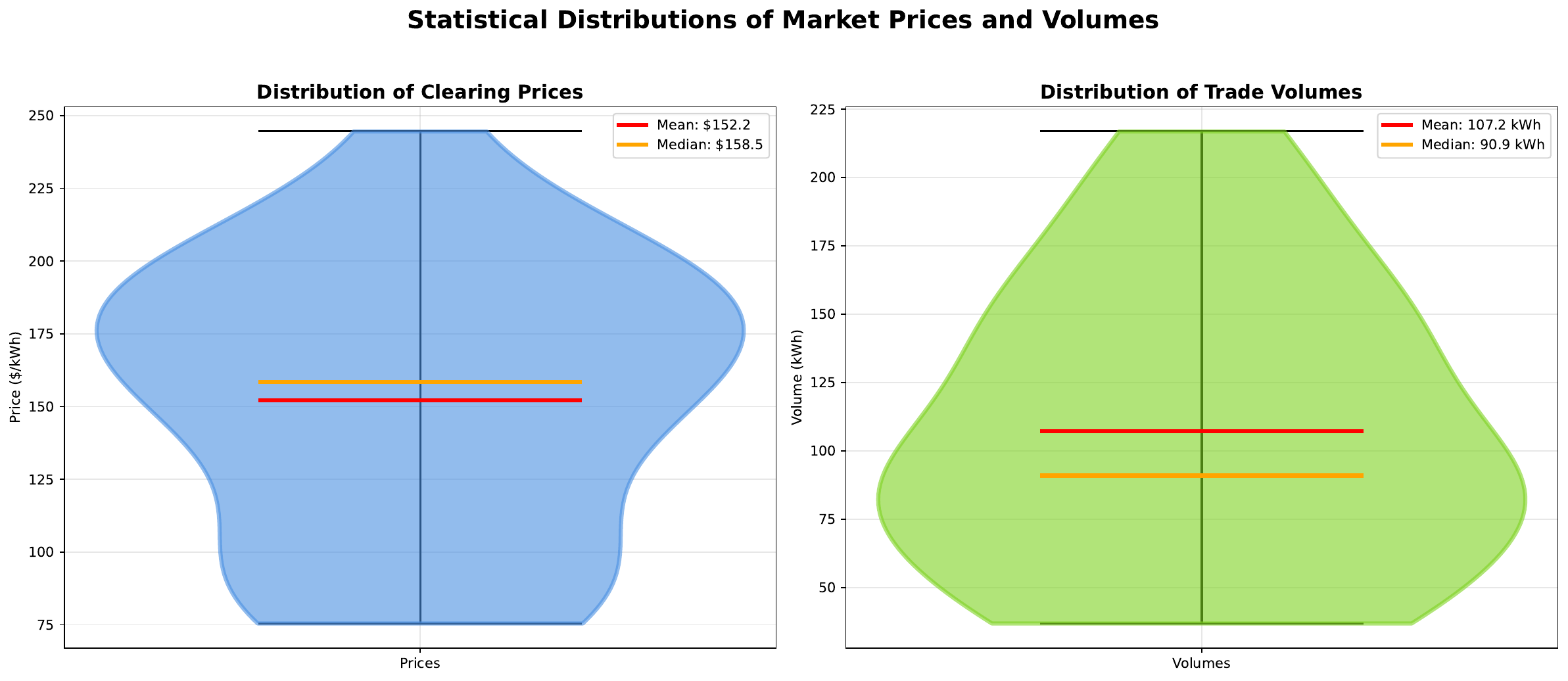}
       \caption{Strategic battery scenario.}
       \label{fig:statistical-distribution-strategic} 
    \end{subfigure}
    \caption{Statistical distributions of market clearing prices and trade volumes for the case study.}
    \label{fig:statistical-distribution}
\end{figure}

A primary function of a battery is temporal arbitrage (buying low and selling high) which stabilizes prices. Our framework captures this economic-physical interaction. Fig. \ref{fig:statistical-distribution} presents a comparative of the statistical distribution of P2P market clearing prices over the 24-hour simulation for both scenarios of the case study.

The \emph{no battery} scenario (see Fig. \ref{fig:statistical-distribution-no-battery}) serves as a baseline and shows a volatile market. The price distribution is multi-modal, with at two distinct price clusters. This is statistical evidence of a fractured market, where agents are entirely subject to the temporal mismatch between generation driving low prices (e.g., the cluster near 140 \$/MWh) and demand driving high prices (e.g., the cluster near 230 \$/MWh). The mean of \$173.6 is pulled away from the median of \$157.5, further indicating a skewed and unstable market.

The \emph{strategic battery} scenario (see Fig. \ref{fig:statistical-distribution-strategic}) demonstrates a transformation of the market structure. The distribution becomes unimodal in a stable price regime that approximates a normal distribution. This is quantitatively confirmed by the fact that the mean of \$152.2 and median of \$158.5 are now almost identical, which is a feature of a symmetrical, stable, and predictable market.

This result demonstrates that the mere presence of strategically configured battery acts as a stabilizing asset, even with non-intelligent agents. The battery provides temporal flexibility, allowing the market to bridge the gap between high-generation and high-demand periods. This arbitrage effectively clips the price extremes and allows the market to converge to a stable clearing price. This finding validates the framework's high-fidelity techno-economic model, proving its capability to capture the emergent, systemic effects that physical assets have on market dynamics.

\subsubsection{Grid stability and system coordination}

\begin{figure}[h]
\centering
\begin{subfigure}{0.49\textwidth}
   \centering
   \includegraphics[width=\linewidth, trim=0cm 13cm 16cm 0cm, clip]{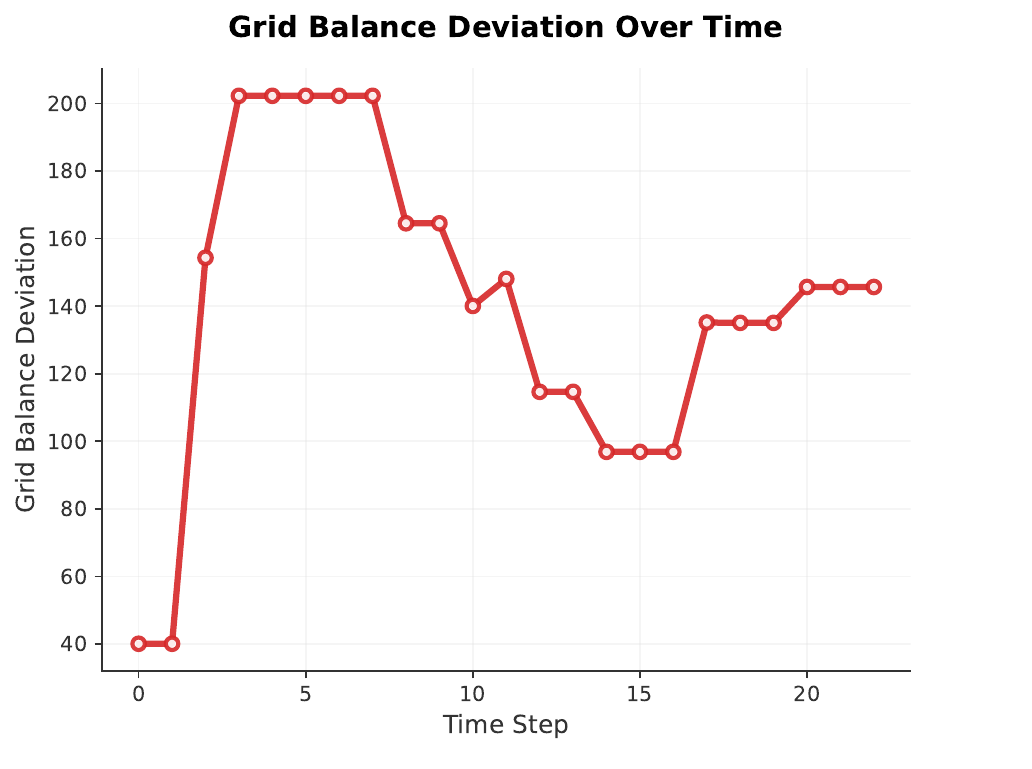}
   \caption{No battery scenario.}
   \label{fig:grid-deviation-no-battery} 
\end{subfigure}
\begin{subfigure}{0.49\textwidth}
   \centering
   \includegraphics[width=\linewidth, trim=0cm 13cm 16cm 0cm, clip]{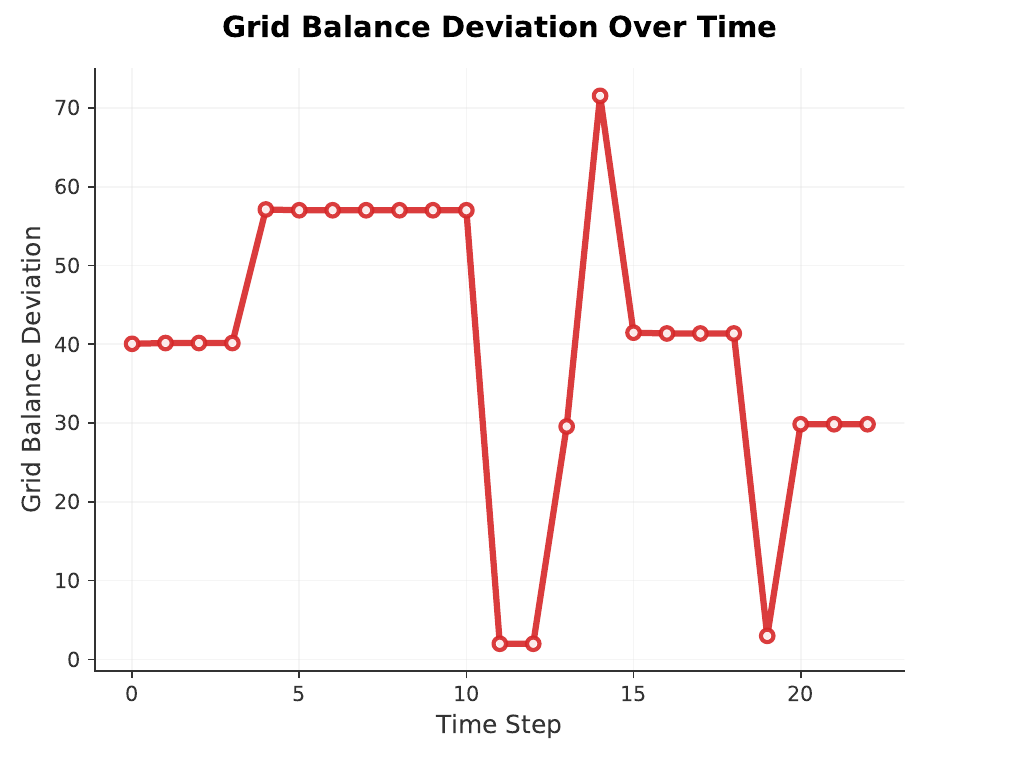}
   \caption{Strategic battery scenario.}
   \label{fig:grid-deviation-strategic} 
\end{subfigure}
\caption{Grid balance deviation over time for the case study.}
\label{fig:grid-deviation}
\end{figure}

\begin{figure}[h]
\centering
\begin{subfigure}{0.49\textwidth}
   \centering
   \includegraphics[width=\linewidth, trim=0cm 0cm 16cm 13cm, clip]{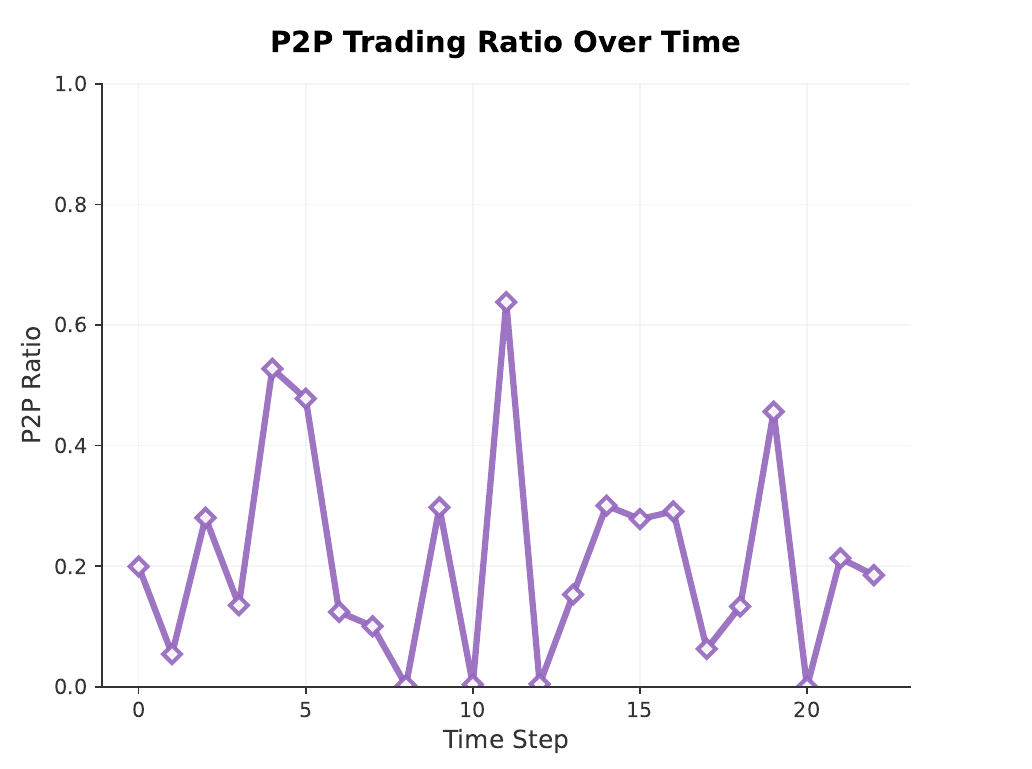}
   \caption{No battery scenario.}
   \label{fig:p2p-ratio-no-battery} 
\end{subfigure}
\begin{subfigure}{0.49\textwidth}
   \centering
   \includegraphics[width=\linewidth, trim=0cm 0cm 16cm 13cm, clip]{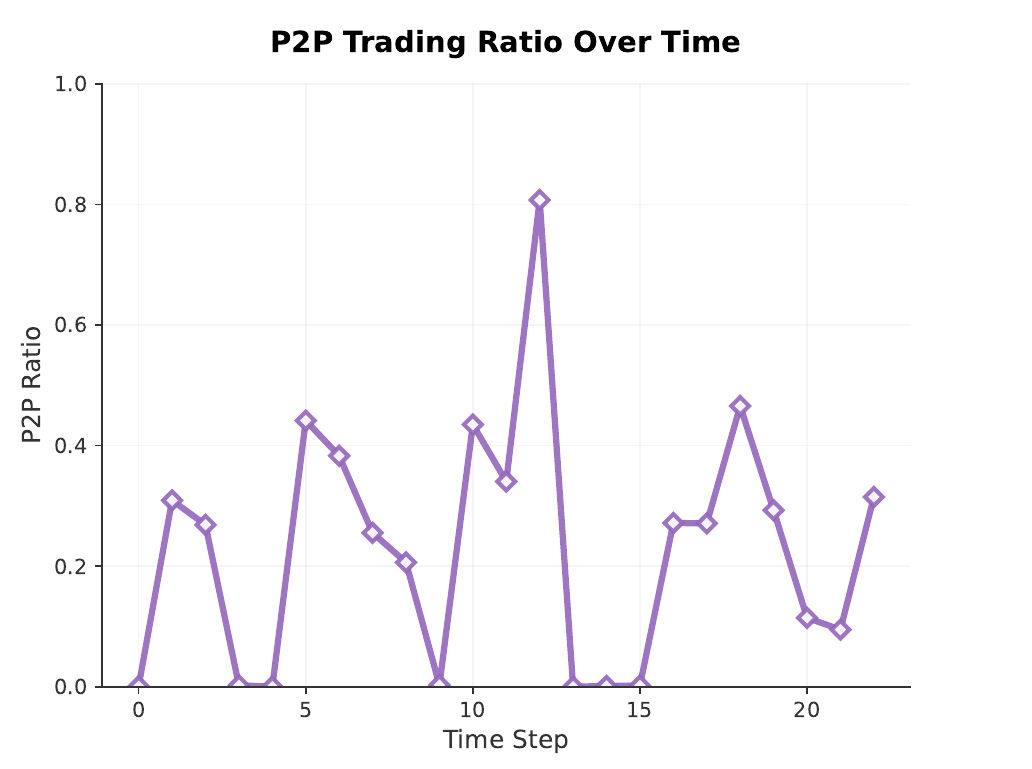}
   \caption{Strategic battery scenario.}
   \label{fig:p2p-ratio-strategic} 
\end{subfigure}

\caption{Peer-to-peer trading ratio over time for the case study.}
\label{fig:p2p-ratio}
\end{figure}

The framework's \emph{Implicit Cooperation Module} is designed to quantify system stability by measuring the supply-demand imbalance and calculating the coordination score as defined in (\ref{eq:coordination}). To illustrate this, Fig. \ref{fig:grid-deviation} and Fig. \ref{fig:p2p-ratio} shows the net grid imbalance and the P2P trading ratio over the 24-hour simulation cycle, respectively.

The \emph{no battery} scenario shows a volatile imbalance profile, mirroring the temporal mismatch between generation and demand. This forces a heavy and continuous reliance on the DSO to act as a sink or source, resulting in a low coordination core.

The \emph{strategic battery} scenario demonstrates a smoother net load profile with a higher overall coordination score. The agents' batteries, even when operated by zero-intelligence policies, provide a passive damper that absorbs systemic disturbances. The design of the \emph{strategic battery} scenario (i.e., agents with morning generation and storage, agents with evening demand and storage) creates a natural, system-wide complementarity.

\subsubsection{Market structure and economic efficiency}

\begin{figure}[h]
\centering
\begin{subfigure}{0.49\textwidth}
   \centering
   \includegraphics[width=\linewidth, trim=0cm 0cm 0cm 1cm, clip]{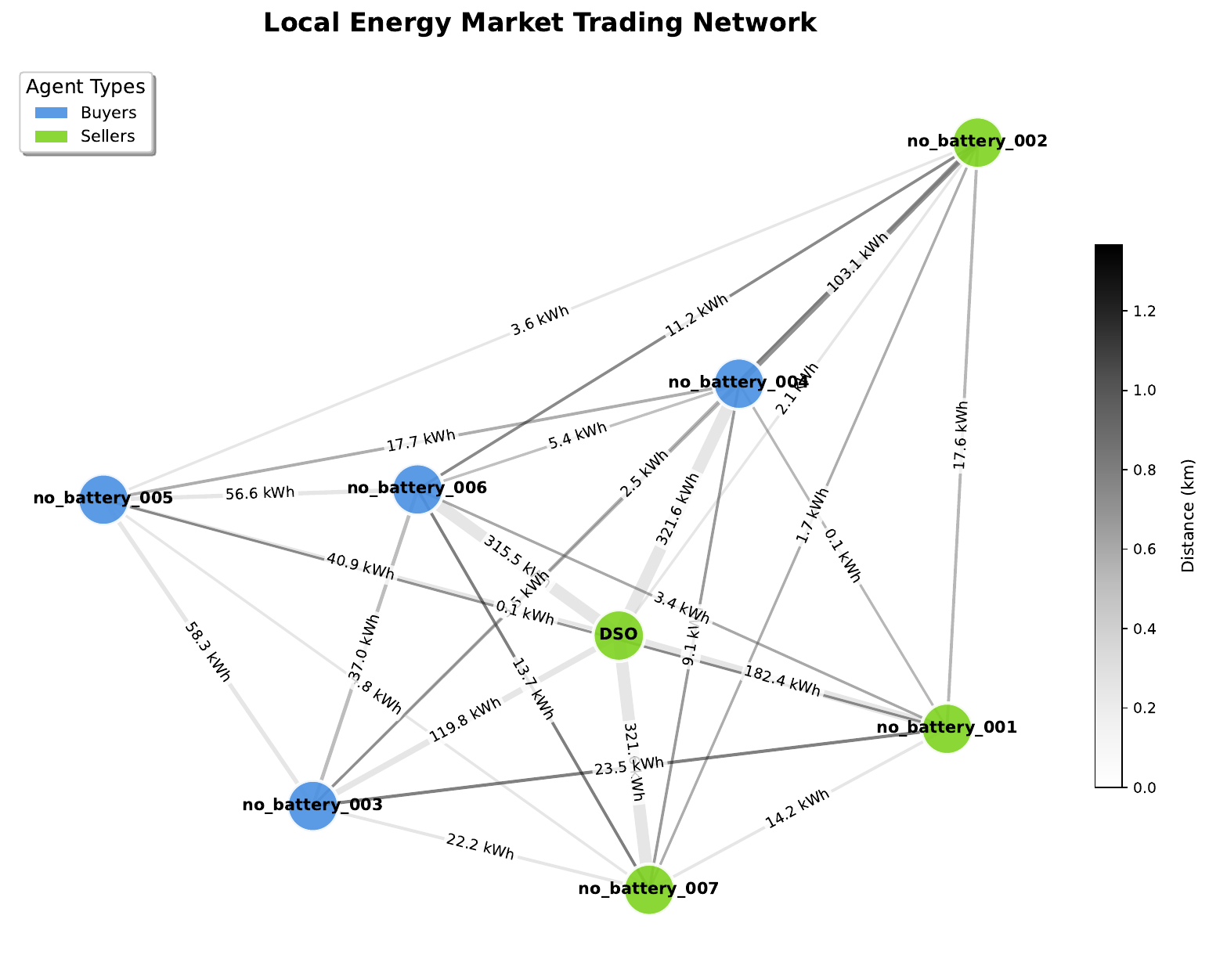}
   \caption{No battery scenario.}
   \label{fig:trading-network-no-battery} 
\end{subfigure}
\begin{subfigure}{0.49\textwidth}
   \centering
   \includegraphics[width=\linewidth, trim=0cm 0cm 0cm 1cm, clip]{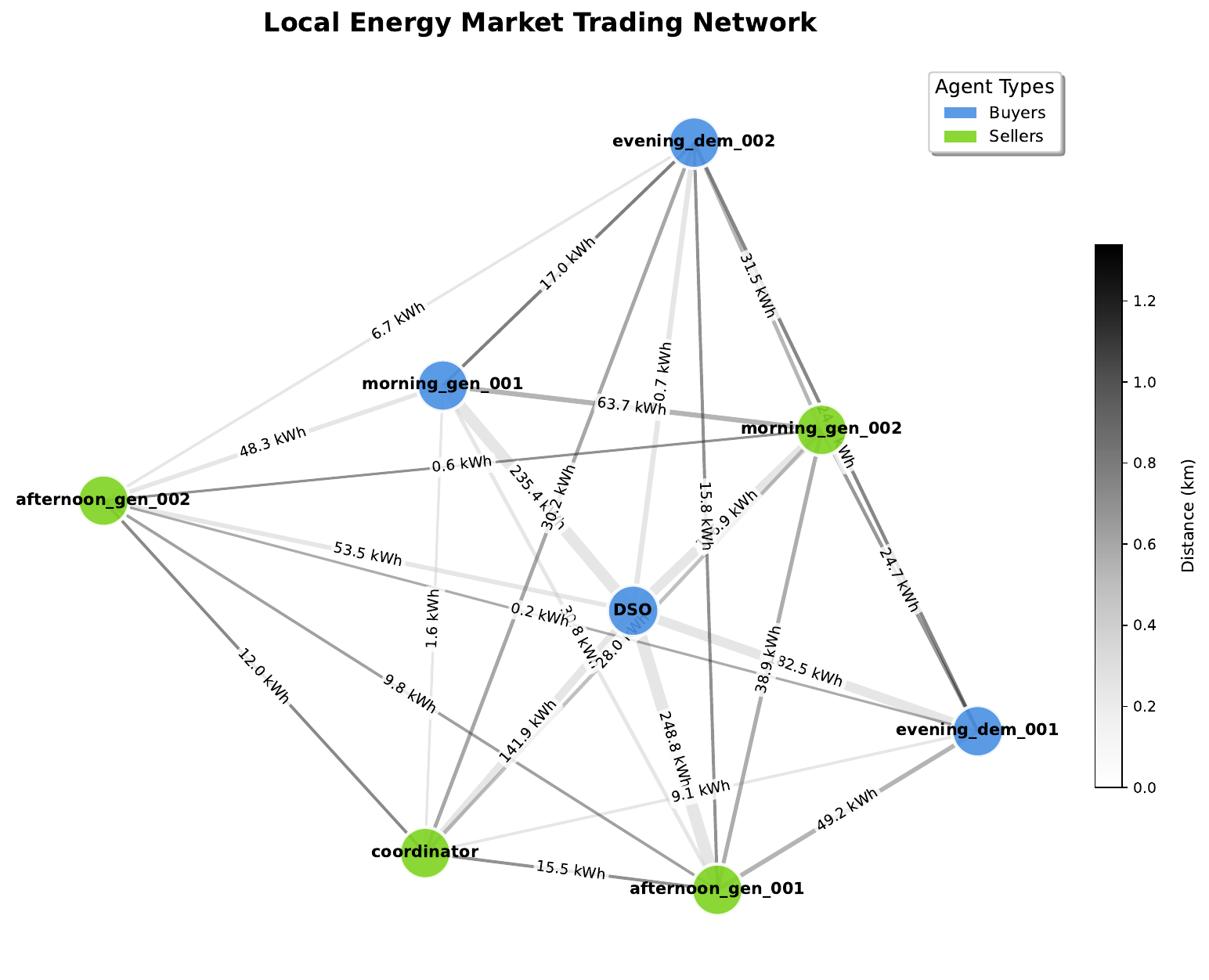}
   \caption{Strategic battery scenario.}
   \label{fig:trading-network-strategic} 
\end{subfigure}
\caption{Local energy market trading network for the strategic battery scenario of the case study.}
\label{fig:trading-network}
\end{figure}

Beyond scalar metrics, the framework's \emph{Visualization Module} provides insights into the structure of the market. Fig. \ref{fig:trading-network} illustrates the structural changes to the market's social and spatial characteristics enabled by storage.

Fig. \ref{fig:trading-network} visualizes the P2P trading network, where nodes represent agents and edges represent the volume of trade between them. The \emph{no battery} scenario (see Fig. \ref{fig:trading-network-no-battery}) depicts a market where most agents (blue nodes) are in a net-demand state, forced to buy all their energy from the DSO (main seller). This demonstrates a market with no liquidity, no self-sufficiency, and a low volume of P2P trading.

The \emph{strategic battery} scenario (see Fig. \ref{fig:trading-network-strategic}) shows that agents have mostly differentiated into clear roles (i.e., agents with morning/afternoon generation are net sellers, agents with evening demand are net buyers), and trade more with each other. This indicates that the temporal flexibility from batteries has created the market. Furthermore, the role of the DSO has reversed from being the main seller in the \emph{no battery} scenario to being the main buyer. This result showcases the LEM has transitioned from a state of energy deficit, dependent on the external grid, to a state of self-sufficiency and is now a net exporter of its surplus. This validation proves the framework's ability to capture systemic changes in market structure and economic flow.

\subsection{Discussion}
\label{1-sec:5.2-discussion}

The results from the zero-intelligence agent simulations validate the framework's design, demonstrating that the underlying techno-economic model is sufficiently high-fidelity to capture the impact of physical asset configurations on market and grid performance. Table \ref{tab:rewards-comparison} presents a quantitative summary comparing the \emph{no battery} and \emph{strategic battery} scenarios, focusing on the aggregate agent rewards, which serve as a proxy for overall market viability and agent profitability.

\begin{table}[]
\centering
\caption{Performance comparison of the scenarios \emph{no battery} and \emph{strategic battery} for the case study.}
\label{tab:rewards-comparison}
\begin{tabular}{@{}lll@{}}
\toprule
\multicolumn{1}{c}{\textbf{Metric}} & \multicolumn{1}{c}{\textbf{No Battery}} & \multicolumn{1}{c}{\textbf{Strategic Battery}} \\ \midrule
Average Reward & -61556.1 & +42929.5 \\
Final Reward & -30827.1 & +78202.6 \\
Battery Ratio & 0.00 & 0.74 \\
Average Efficiency & - & 95\% \\
SOC Range & - & 10-90\% \\ \bottomrule
\end{tabular}
\end{table}

The \emph{strategic battery} scenario achieves a positive average aggregate reward of +42929.5, a reversal from the baseline's large negative reward of -61556.1. This demonstrates the economic and systemic value that strategically deployed storage resources, even when operated non-strategically, provide to the market ecosystem by enabling temporal arbitrage. This validation provides a testbed for quantifying the value of different system designs and physical assets.

The framework is proven to be sensitive to the physical prerequisites for coordination (i.e., storage); the \emph{strategic battery} scenario, for instance, creates clear, complementary roles that are ideal for future MARL agents to learn to fill. While this study held the market mechanism constant, the influence of storage on price volatility directly demonstrates the framework's capability to analyze the coupling between market mechanisms and physical assets. Furthermore, this case study directly addresses the impact of physical grid constraints. The battery parameters (e.g., SoC or efficiency) are a form of temporal physical constraint, and their presence influences agent behavior and market outcomes, validating the framework's capacity to test the market-versus-grid dichotomy.

While the presented 7-agent study validates the framework's functional capabilities, a formal quantitative scalability analysis remains a critical component of future work. This would involve executing simulations with increasing agent populations (e.g., 10, 50, 100+ agents) to plot computational time and analyze the impact of agent count on market KPIs, thereby testing the framework's performance under load.

We must reiterate that the results presented in this paper are illustrative and employ a zero-intelligence agent baseline to validate the simulation framework's capability. A full MARL training is required to understand the emergent behavioral dynamics and the full potential of implicit cooperation (authors will publish this work in a forthcoming paper).

Finally, as with all simulations, the component models are abstractions that require continuous validation and calibration against real-world data to improve their fidelity.

\section{Conclusion and future work}
\label{1-sec:6-conclusion}

In this paper, we have presented a novel, open-source simulation framework specifically designed for the study of decentralized LEMs using MARL. Addressing a gap in the current landscape of energy simulation tools, specifically the lack of a unified platform that integrates realistic market mechanisms, physical grid constraints, and advanced MARL capabilities within a standardized and decentralized paradigm, our framework provides a robust, extensible, and Gymnasium-compliant environment for researching the complex dynamics of future energy systems.

The primary contribution is the LEM simulation framework itself. Its architecture and formulation as a Dec-POMDP overcomes the prevalent market-versus-grid dichotomy by cohesively unifying a modular market platform (featuring a hybrid preference/price-based matching and reputation systems) with a physically grounded grid model (incorporating transmission losses and congestion management). The framework is architected to support and facilitate research into fully decentralized learning and execution paradigms, moving beyond the limitations of centralized training assumptions common in prior work.

A second key contribution is the framework's explicit focus on enabling the study and fostering of implicit cooperation. We achieve this through a novel methodological approach embedded within the framework: the enrichment of agents' partial observations with system-level KPIs and the integration of these KPIs into a multi-objective reward function. This creates a unique feedback loop, demonstrating a viable pathway for guiding autonomous, self-interested agents towards collectively beneficial, coordinated behaviors without necessitating explicit communication or centralized control.

This work lays the foundation for future research enabled by the developed framework. The immediate next step is a comprehensive validation of the implicit cooperation mechanisms facilitated by this framework. This involves conducting large-scale MARL training experiments utilizing the framework's full capabilities. Specifically, we will train agents using the implemented state-of-the-art MARL algorithms (i.e., PPO, APPO, and SAC) across the spectrum of training and inference paradigms: CTCE as a benchmark, the practical CTDE approach, and, most importantly, the fully decentralized DTDE paradigm that aligns with real-world LEM principles. These experiments will systematically evaluate how different market designs, grid conditions, and learning parameters influence the emergence and effectiveness of implicit coordination, quantifying the trade-offs between decentralization, performance, and stability.

Beyond this core validation, the modularity and extensibility of the framework open several promising avenues for future research:

\begin{itemize}
    \item \textbf{Integrating more sophisticated DER models:} Enhancing the fidelity of agent models by incorporating more detailed representations of battery degradation, charging/discharging patterns of electric vehicles, building thermal dynamics, or heterogenous agent objectives (e.g., risk aversion, fairness concerns).

    \item \textbf{Exploring advanced MARL algorithms:} Implementing and benchmarking a wider suite of cutting-edge MARL algorithms, including those incorporating communication protocols (to contrast with implicit methods), alternative value decomposition techniques, or model-based approaches.

    \item \textbf{Scalability analysis:} Conducting a formal quantitative scalability analysis by executing simulations with increasing agent populations (e.g., 10, 50, 100+ agents) to plot computational time and analyze the impact of agent count on market KPIs.

    \item \textbf{Application to real-world data:} Validating and calibrating the simulation framework using empirical data from pilot LEM projects or real distribution networks to enhance its accuracy and practical relevance for policy analysis.

    \item \textbf{Analysis of communication constraints:} Extending the framework to explicitly model the impact of realistic communication network limitations (latency, bandwidth) and potential threats (e.g., false data injection) on the robustness and security of decentralized market operations.
\end{itemize}

We believe this LEM simulation framework will serve as a useful tool for the research community, policymakers, and industry practitioners striving to design, analyze, and deploy the intelligent, resilient, efficient, and truly decentralized energy systems required for a sustainable future.

\bibliographystyle{elsarticle-num}
\bibliography{Literature}

\end{document}